\documentclass[fleqn,usenatbib]{mnras}
\pdfoutput=1 
\usepackage{newtxtext,newtxmath}

\usepackage[T1]{fontenc}

\DeclareRobustCommand{\VAN}[3]{#2}
\let\VANthebibliography\thebibliography
\def\thebibliography{\DeclareRobustCommand{\VAN}[3]{##3}\VANthebibliography}

\usepackage{graphicx}	% Including figure files
\usepackage{amsmath}	% Advanced maths commands
\usepackage{amssymb}	% Extra maths symbols
\usepackage[switch, modulo]{lineno}

\title[Dynamics of dust in the coma of 67P]{Distribution and dynamics of decimeter-sized dust agglomerates in the coma of 67P/Churyumov--Gerasimenko.}

\author[P. Lemos]{
Pablo Lemos,$^{1,2}$\thanks{E-mail: j.lemos-velazquez@tu-braunschweig.de}, Jessica Agarwal$^{1,2}$, Matthias Schr\"oter$^3$
\\
$^{1}$Institut f\"ur Geophysik und Extraterrestrische Physik, Technische Universit\"at Braunschweig, Mendelssohnstraße 3, Braunschweig 38106, Germany.\\
$^{2}$Max Planck Institute for Solar System Research, Justus-von-Liebig-Weg 3, G\"ottingen 37077, Germany.\\
$^{3}$Max Planck Institute for Dynamics and Self-Organization, Am Faßberg 17, D-37077 Göttingen, Germany.
}

\date{Accepted XXX. Received YYY; in original form ZZZ}

\pubyear{2022}

\begin{document}
\label{firstpage}
\pagerange{\pageref{firstpage}--\pageref{lastpage}}
\maketitle

\begin{abstract}
We present a method to analyze images of the coma of 67P/Churyumov–Gerasimenko obtained using OSIRIS, the main imaging system onboard \textit{Rosetta}, where dust aggregates can be seen as bright tracks because of their relative velocity with respect to the spacecraft. We applied this method to 105 images taken in 2015 July, 2015 December and 2016 January, identifying more than 20000 individual objects. We performed a photometric analysis of them, finding their phase function. This phase function follows the same trend as the one found for the nucleus, consistent with the detected particles having a size larger than $\sim 1$ mm. Additionally, the phase function becomes shallower for increasing heliocentric distances, indicating a decrease in the mean agglomerate size. In order to characterize the agglomerates observed in the image, we developed a simplified model for their ejection and dynamics in the coma, and generated synthetic images based on it. We solved the inverse problem by finding the simulation parameters that give the best fit between synthetic and real images. In doing so, we were able to obtain a mean agglomerate size $\sim$ dm and initial speed $\simeq$ 1 m s\textsuperscript{-1}. Both show a decrease with increasing heliocentric distance, sign of the reduction in activity. Also, the sizes obtained by the comparison are not compatible with ejection caused by water activity, so other sources have to be invoked, mainly CO\textsubscript{2}.
\end{abstract}


%%%%%%%%%%%%%%%%%%%%%%%%%%%%%%%%%%%%%%%%%%%%%%%%%%

%%%%%%%%%%%%%%%%% BODY OF PAPER %%%%%%%%%%%%%%%%%%

\section{Introduction}

The \textit{Rosetta} mission provided data with unprecedented detail on comet 67P/Churyumov--Gerasimenko (hereafter 67P) by sampling its environment \textit{in situ} during a period of around two years. In particular, cometary dust particles over a wide range of sizes were collected, analyzed and characterized by MIDAS (for particles in the range $\mu$m to tens of $\mu$m, \citealt{Mannel2019,Longobardo2022}), COSIMA (tens of $\mu$m -- hundreds of $\mu$m, \citealt{Merouane2017}) and GIADA (hundreds of nm -- tens of mm, \citealt{DellaCorte2019,Longobardo2022}) instruments. Larger objects ($\gtrsim$ 1 cm) could be detected by the main imaging system onboard Rosetta, the Optical, Spectroscopic and Infrared Remote Imaging System (OSIRIS, \citealt{Keller2007}). The data obtained by OSIRIS make it possible to obtain information about the morphological and dynamical properties of the dust, and in case the same object could be identified in more than one image while using different filters, also about its color, which can give hints about its composition \citep{Frattin2017,Kwon2022}. 

However, remotely analyzing individual dust particles or aggregates in the coma must face a fundamental issue: the distance from the sensor to the object is unknown, so its size and velocity, and hence size and mass distribution, cannot be uniquely determined. Several works deal with this issue in different ways: \citet{Rotundi2015} and \citet{Fulle2016} assume that the motion of the objects is entirely radial from the nucleus, and that the apparent motion with respect to the camera comes mainly from the spacecraft velocity. Using those assumptions, the distance can be determined using the parallax effect. \citet{Agarwal2016} and \citet{Pfeifer2022} use images where the nucleus limb is present, and focus on agglomerates going away from it. These agglomerates have a higher probability of being recently ejected, so it can be assumed that they are at the same distance as the nucleus; \citet{Drolshagen2017} and \citet{Ott2017} exploit the fact that the two detectors of OSIRIS, the Narrow (NAC) and Wide (WAC) Angle Cameras, are separated by $\simeq70$ cm on the spacecraft, so if both cameras detect the same object, the parallax effect can be used to measure its distance to the camera; in \citet{Guttler2017} it is noted that objects closer than $\sim100$ m appear unfocused in WAC images, so the authors develop a method to measure the distance to objects close to the camera by measuring the apparent size of the unfocused pattern, which is directly related to its distance; finally, \citet{Frattin2021} uses a mixed approach, constraining the sizes and distances of the dust agglomerates based on speed distributions taken from some of the works listed before, in combination with photometric simulations. 

The approach of this work is different from that of its predecessors. Instead of looking for an alternative method for determining the distance, we propose to bypass this requirement by using a combination of observations and statistical modelling. On the one hand, images taken by OSIRIS are analyzed in order to obtain a set of observables from the distribution of dust agglomerates present in each of them. On the other hand, we simulate the trajectories of dust agglomerates through the coma using a simplified ejection and dynamical model. These trajectories are characterized by different dust parameters, such as size, density and initial velocity. Based on these simulations and the spacecraft position and orientation, a group of synthetic images of the agglomerates as seen by OSIRIS are generated. Using these synthetic images, the inverse problem is solved by optimizing the parameter choice for the dynamical simulations in order to reproduce properties of the dust agglomerate trajectories observed in the real images. This approach is also different from that applied in previous works in that the properties of the entire population of detected objects are analyzed statistically, rather than dealing with individual objects as done previously. 

The work is organized as follows: in Section \ref{sec:observations} the datasets and the dust agglomerate detection method are described. The dynamical model and the synthetic image generation are explained in Section \ref{sec:model}. In Section \ref{sec:realtracks} we present the analysis of the properties of dust agglomerates found in OSIRIS images. In Section \ref{sec:comparison} the synthetic images are compared with the real ones. Finally, we present our conclusions in Section \ref{sec:conclusion}.

\section{Observations and tracks detection}\label{sec:observations}

\begin{figure}
    \centering
    \includegraphics[width = \linewidth]{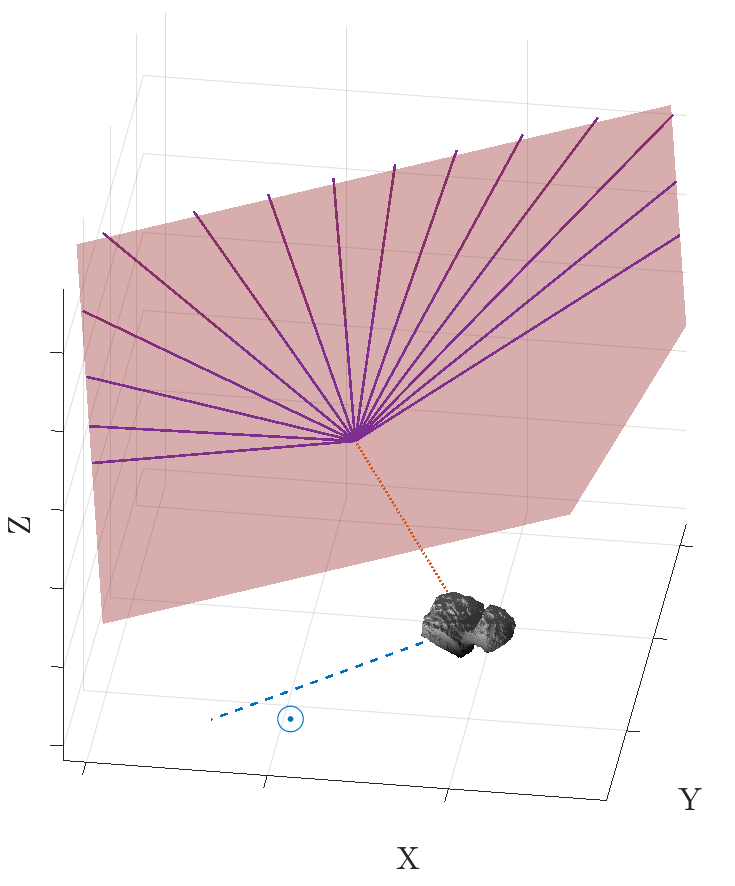}
    \caption{Sketch of the observation geometry. The solar (blue dashed) and spacecraft (red dotted) directions with origin in the nucleus form a perpendicular angle. The pointing of the camera, shown as violet solid lines, scan the coma for different phase angles in the plane roughly perpendicular to that of the spacecraft, nucleus and Sun. Note that the image is not to scale.} 
    \label{fig:sketch}
\end{figure}

\textit{Rosetta} escorted 67P from 2014 August when its heliocentric distance was $\simeq 3.7$ au inbound, to 2016 September when it was outbound at $\simeq 3.8$ au from the Sun. For this work we will focus on three different image sets obtained with the OSIRIS NAC around perihelion. All these image sequences were obtained under the operational activity DUST\_PHASE\_FUNCTION, originally devoted to the analysis of the dusty coma brightness as a function of the phase angle (i.e. the angle between the Sun-spacecraft and camera pointing direction) and the wavelength. In order to achieve this, the observing conditions were such that the distance from the nucleus to the spacecraft remained nearly constant throughout the duration of the acquisition, while the camera pointing scanned the coma at different phase angles. The plane of observation was nearly perpendicular to that containing the Sun, the nucleus, and the spacecraft. A sketch of the observation geometry can be seen in Fig. \ref{fig:sketch}. 

The image sets used in this work were acquired on 2015 July 7, 2015 December 14 and 2016 January 21, at heliocentric distances of 1.32 au inbound, 1.89 au outbound and 2.18 au outbound respectively. All three sets were taken using the Blue F24 (peak transmission at 480.7 nm), Orange F22 (649.2 nm) and Red F28 (743.7 nm) filters. A binning of $4\times4$ was used for all images, so the final image size is $512\times512$ pixels. A summary of the observing conditions can be found in Table \ref{tab:obs}.

\begin{table*}
    \centering
    \caption{Image sets used for this work. Columns represent the mid and short-term planning cycles (intervals of roughly one month and one week), date of acquisition, filters and exposure times used, heliocentric distances, nucleocentric distances and number of images in the set.}
    \begin{tabular}{c|c|c|c|c|c|}
        \hline
        Planning cycle (MTP/STP) & Date & F ($t_{exp}$) & $r_h$ (au) & $r_{S/C}$ (km) & $\#I$\\
        \hline
        018/063 & 2015-07-07 & F22 (7 s), F24 (73 s), F28 (40 s) & 1.32 & 153.4 & 45\\
        023/086 & 2015-12-14 & F24 (73 s), F22 (7 s), F28 (40 s) & 1.89 & 102.6 & 21\\
        025/092 & 2016-01-21 & F24 (146 s), F22 (14 s), F28 (80 s) & 2.18 & 79.2 & 39\\
        \hline
    \end{tabular}
    \label{tab:obs}
\end{table*}

Depending on the heliocentric distance and the filter used, the images were obtained using exposure times ranging from 7 to 146 seconds. These exposure times combined with the nonzero relative velocity between the spacecraft and the dust agglomerates result in them appearing in the images not as point sources, but instead as elongated tracks. This fact will be exploited later, at the moment of the object detection.

A total of 105 level 3F images were used for this work. These images are radiometrically calibrated, corrected for geometric distortion and for solar and in-field stray light, and expressed in reflectance units, i.e. the corrected flux is normalized by the solar flux at the corresponding heliocentric distance. A detailed description of the data processing steps can be found in \citet{Tubiana2015}. Despite being corrected for stray light effects, some of the high phase angle images present illumination artefacts that complicate the track detection. This problem is more evident in images taken at phase angles greater than 100\degr, that is, when the camera pointing is closer to the Sun direction, so the results in this range should be treated with caution.

\subsection{Detection method}\label{sec:detMethod}
\begin{figure}
    \centering
    \includegraphics[width = .9\linewidth]{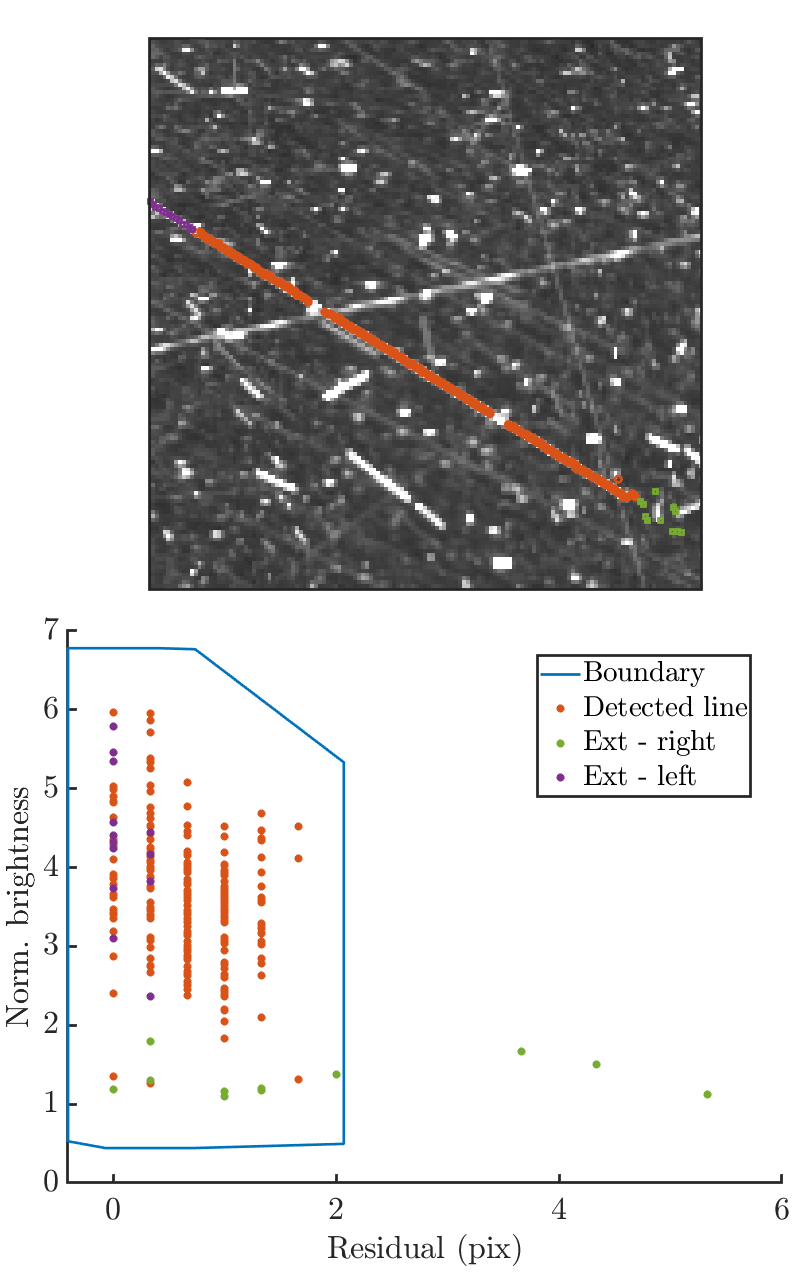}
    \caption{Example of the track correction and extension method. Top: The orange points represent the peak values positions from the segments perpendicular to the track, which was obtained from the Hough transform. The detected track fails to cover the entire length on the left side (the gaps in between are caused by overlapping background stars). Using these positions and brightness values, the boundary region is defined in the bottom panel. The same procedure is made for profiles in the extended region. The violet symbols from the extended track to the left lie inside this boundary region, so the line is extended. On the contrary, the green points on the right side are not.}
    \label{fig:ext}
\end{figure}

A semi-automatic detection method based on the one presented in \citet{Frattin2017} was used. The steps involved in this method are:

\begin{itemize}
    \item A similarity map $SM_{\theta}$ is created using a track template $T_{\theta}$. These templates consist of a square window of 10 pixels in length, where a straight line representing the track passes through the centre of the template. The orientation angle $\theta$, defined as $\theta=\arctan(-1/m)$\footnote{We use this definition for the orientation angle in order to match the one used in the Hough transform later in the algorithm.}, where $m$ is the slope of the line, successively takes all the values in the $[-90\degr,+89\degr]$ range, with steps of $4\degr$. $SM_{\theta}$ is calculated as the normalized cross correlation (NCC) between each image $I$ and the template $T_{\theta}$, and applying the convolution of the result with the same template
\begin{equation}
    SM_{\theta} = \left( I \Bar{\otimes} T_{\theta}\right) \otimes T_{\theta}.
\end{equation}

    \item Binary images are generated from the similarity maps for each orientation by imposing a lower threshold defined as $J+2S$, where $J$ and $S$ represent the local median and standard deviation of the NCC respectively. Nonzero pixels in these binary images represent locations in the image with high probability of having a track with a determined orientation. 
    
    \item Tracks are detected from each binary image using a Hough transform method (\citealt{hough1962method}, \citealt{Duda1972}). The outcome of this step is called \textit{nominal track}.
    
    \item To characterize the nominal tracks, segments perpendicular to the track are analyzed. The centres of the segments are equally spaced on the track, with a distance of 1/3 pixel between them. Brightness profiles are then generated by interpolating the image values over the segment positions. For each profile, two parameters are defined: its brightness peak value, and the residual distance to the nominal track, defined as the distance in pixels from the nominal track to the peak position along the mentioned segment. Once these parameter pairs are defined for all segments, the track is characterized by a \textit{boundary region}, i.e. a region in the brightness--residual space enclosed by the convex hull of all the pairs, extended by the standard deviation along each axis (Fig. \ref{fig:ext}).
    
    \item The nominal tracks are corrected for incomplete detection. First, the nominal track is preliminarily extended by 5 pixels. Then, the brightness--residual pairs are defined for the extended part, and compared with the boundary region defined in the previous step. If the points corresponding to the extended part lie inside the boundary region, the line is extended. The process is repeated until the added points do not belong to the region or the image edge is reached. An example of this process is shown in Fig. \ref{fig:ext}. 
    
    \item The extended tracks are analyzed in the search for duplicate detections. This is done by comparing the pixels spanned by the tracks. If two extended tracks share more than 70\% of their pixels, the tracks are merged. 
    
    \item A manual inspection and correction of the results is performed.
\end{itemize}

By using this method, a total of 20033 tracks were detected. This number is larger by at least an order of magnitude from previous studies focused in detecting and analyzing this type of tracks in similar images. It is worth noticing that since the track templates $T_{\theta}$ have a size of 10 pixels, our algorithm is unable to find tracks shorter than that length. A smaller template size would have meant that shorter tracks could be detected, but also increases the chance of mistakenly identifying a group of bright background pixels as a real track. In order to check if this choice of template size introduces a bias in the detected tracks, we checked the properties a dust agglomerate must have to generate a track of this length. The projected track length in pixels $l_{pix}$ depends on the agglomerate projected speed $v$, distance to the camera $d$ and the image exposure time $t_{exp}$. The equation describing the track length as a function of these variables is 

\begin{equation}
    l_{pix} = \frac{t_{exp}}{d\,R_{NAC}} v,
\end{equation}

\noindent where $R_{NAC}$ is the angular resolution of the camera. For image sets similar to the ones used here, \citet{Frattin2021} estimated the maximum agglomerate to camera distance of 18 km. At that distance, the minimum agglomerate projected speed needed for generating a track longer than 10 pixels is 0.4 m s\textsuperscript{-1} for images taken with the Blue and Red filters, but 1m s\textsuperscript{-1} for the ones corresponding to the Orange filter. Since \citet{Ott2017} found that the median apparent speed of this type of agglomerates is 0.6 m s\textsuperscript{-1}, we can conclude that the tracks detected in the images taken with the Orange filter sample a population of agglomerates that has a higher relative speed to the spacecraft, is closer to the camera, or a combination of both. 

\section{Image modelling}\label{sec:model}

\subsection{Dust dynamics model}\label{sec:dustdyn}

Synthetic images were generated by modelling the trajectories of different types of dust agglomerates in the comet coma, and looking for intersections with the camera FOV. A simplified model for computing the dust agglomerate trajectories was developed. This model is initially developed in 2D, and assuming the activity is axially symmetrical with respect to the solar direction, the 3D trajectories are obtained by rotating the 2D ones with respect to the solar direction by a random angle. In this model, the nucleus is represented by a sphere of radius $R_N = 2000$ m and mass $M_N = 9.982\times 10^{12}$ kg. The dust agglomerates are assumed to be spherical with radius $r_d$ and density $\rho_d$, and are under the influence of three forces: nucleus gravity $F_G$, radiation pressure $F_R$ and gaseous drag $F_D$, expressed as
\begin{align}
    \mathbf{F_G} & = - \frac{\mathcal{G} M_N m}{r^2} \frac{\mathbf{r}}{r}\label{eq:system_1}\\
    \mathbf{F_R} & = - \frac{c_{\odot} Q_{RP} \pi r_d^2}{r_h^2 c} \frac{\mathbf{r_h}}{r_h}\\
    \mathbf{F_D} & = \frac{|\mathbf{v_g} - \mathbf{v_d}|^2}{2} \rho_g \pi r_d^2 C_D \frac{\mathbf{V}}{V}\label{eq:system_3},
\end{align}

\noindent where $m=4/3 \pi \rho_d r_d^3$ is the object mass, $\mathbf{r}$ is the position of the object with respect to the nucleus, $\mathbf{V}=\mathbf{v_g} - \mathbf{v_d}$ is the relative velocity between the agglomerate and the gas, $c_{\odot} = 1361$ W m\textsuperscript{-2} is the solar constant, $Q_{RP}$ is the scatter efficiency for radiation pressure (assumed to be equal to 1), $r_h$ is the heliocentric distance expressed in au, $c$ is the speed of light, $\mathbf{v_g}$ and $\rho_g$ are the gas velocity and density respectively, and $C_D$ is the drag parameter, calculated using the free-molecular expression \citep{1994mgdd.book.....B} as

\begin{equation}
    C_D = \frac{2 s^2 +1}{s^3 \sqrt{\pi}} \exp{(-s^2)} + \frac{4s^4 + 4s^2-1}{2s^4} \text{erf}(s) + \frac{2\sqrt{\pi}}{3s} \sqrt{\frac{T_d}{T_g}},
\end{equation}

\noindent where $s=V/\sqrt{2T_g k_B/m_g}$, and the dust temperature $T_d$ is assumed to be equal to the gas one $T_g$. Computing the gas drag force requires a description of the density and velocity of the gas in the coma, for which an intermediate step needs to be included (see Section \ref{sec:gasmodel}).

The initial position of the dust agglomerates is chosen from a probability distribution function that has the same dependence on the subsolar angle as the gas production rate, obtained from the model by \citet{Fulle2020} (see Sec. \ref{sec:gasmodel}). The initial velocity modules are chosen from a Maxwell--Boltzmann distribution 

\begin{equation}\label{eq:maxw}
    f(v)=\frac{4}{\sqrt{\pi}} \frac{v^2}{{v_P}^3} \exp{-(v^2/{v_P}^2)},
\end{equation}
\noindent where $v_P$ is the most probable speed. In order to represent the surface roughness in a simplified way, we include a tangential component to the initial velocity, such as its direction forms an angle $\theta_i$ with the local normal. $\theta_i$ is chosen from a normal distribution centred on the free parameter $\theta_d$ and with standard deviation of $20\degr$, except for the case $\theta_i=0\degr$, when all the agglomerates start with radial velocities.

The dust agglomerates are then characterized by four parameters: their density $\rho_d$, radius $r_d$, most probable initial speed $v_P$ and most probable initial direction $\theta_d$ from the surface normal. The dynamical simulations were carried out individually for each combination of those parameters. The values used for each parameter are listed on Table \ref{tab:dustprop}. A total of 1176 parameter combinations were used for the dynamical integrations, except in the case of STP063, where 1470 combinations were used. Even though some of these combinations do not represent any physically meaningful particle, they are none the less simulated in order to better comprehend the impact of the parameter choice on the results.

\begin{table}
    \centering
    \caption{Values used for the dynamical simulations for each dust parameter. The radii marked with a \textsuperscript{*} were only used for the set STP063.}
    \begin{tabular}{c|c}
        \hline
        Parameter & Values \\\hline
        $\rho_d$ & $[1;10;50;100;200;500;800]$ kg m\textsuperscript{-3}\\
        $r_d$ & $[0.01; 0.05; 0.1; 0.5; 1; 5; 10; 50; 80^*; 100^*]$ cm\\
        $v_P$ & $[0;0.5;1.0;2.0;5.0;10.0]$ m s\textsuperscript{-1}\\
        $\theta_d$ & $[0;20;40;60]\;\degr$ \\
        \hline
    \end{tabular}
    \label{tab:dustprop}
\end{table}

\subsection{Gas model} \label{sec:gasmodel}

\begin{figure}
    \centering
    \includegraphics[width = .8\linewidth]{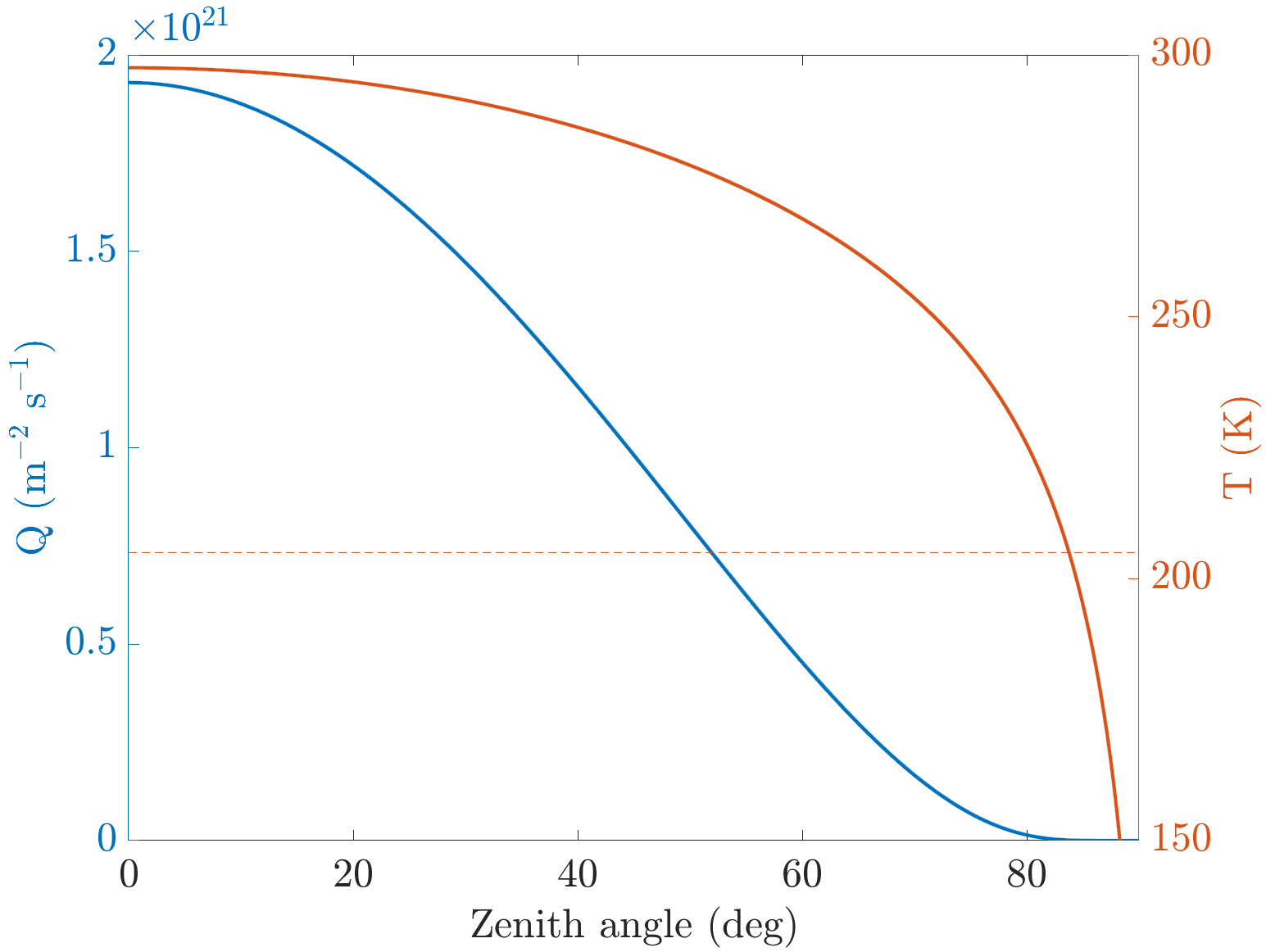}
    \includegraphics[width = \linewidth]{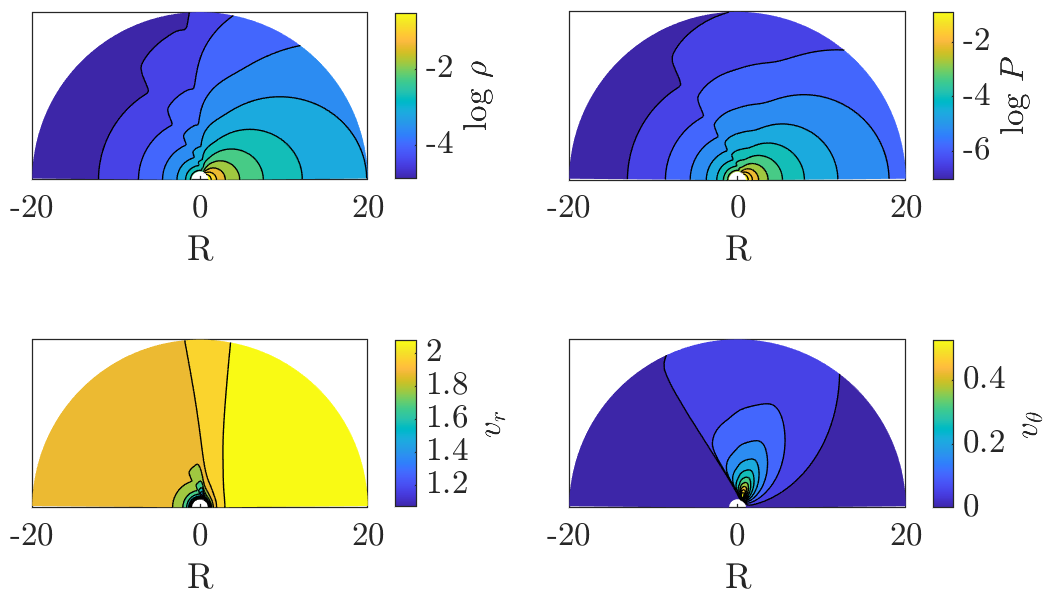}
    \caption{Gas simulations for the set STP092. Top: production rate per area unit and surface temperature as a function of the insolation angle. The dashed line indicates the 205 K limit from which dust ejection is possible. Bottom: Static solution for the gas flow. From top left to bottom right, the panels represent density, pressure, radial and tangential velocities in arbitrary units. The nucleus is at the origin, the illumination comes from the positive $x$ axis and the distances are expressed in units of $R_N$.}
    \label{fig:gasdist}
\end{figure}

The gas simulations are done in two parts. First, the gas production rate is calculated based on the model presented by \citet{Fulle2020}. This model assumes the nucleus surface to be composed of cm-sized pebbles and water ice sublimating inside them. When the surface temperature is larger than 205 K, the pressure inside the pebble is high enough to overcome its tensile strength, making dust ejection possible. Using the heliocentric distances obtained from the header of the images, the production rate as a function of the subsolar angle is calculated (Fig. \ref{fig:gasdist}). 

For the second part, this production rate is used as a boundary condition for the hydrodynamic simulations for the distribution of gas in the coma. As in previous works \citep{Zakharov2018,Zakharov2021}, the initial speed of the gas on the nucleus surface is set to the local sound speed. The gas flow is modelled through the Euler equations, which imply the gas is considered to be ideal, at equilibrium and without viscous dissipation or heat conductivity. The hydrodynamic simulations are carried out in 2 dimensions using the code PLUTO \citep{Mignone2007} until a static solution is achieved (Fig. \ref{fig:gasdist}).

\subsection{Generation of synthetic images}\label{sec:synthIm}

Using the results of the gas simulations discussed in Section \ref{sec:gasmodel}, the system described by equations \ref{eq:system_1}--\ref{eq:system_3} can be numerically solved. The two-dimensional dust agglomerate trajectories obtained from the dynamical modelling are then transformed into three dimensions by using the symmetry assumptions mentioned in Section \ref{sec:dustdyn}. These three-dimensional trajectories are checked for possible intersections with the camera FOV. If such intersection occurs, two intersection points, entry and exit, are defined. A random position $\mathbf{r}_1$ inside the FOV is selected from a linear interpolation between the intersection points. This will be used as one of the endpoints of the synthetic track. The remaining endpoint is defined as $\mathbf{r}_2 = \mathbf{r}_1 \pm \mathbf{v}_1 \times t_{exp}$, where $\mathbf{v}_1$ is the interpolated velocity at $\mathbf{r_1}$, $t_{exp}$ is the image exposure time and the sign is chosen randomly. While $\mathbf{r}_1$ is enclosed within the camera FOV, that is not necessarily the case for $\mathbf{r}_2$. Both endpoints are then projected into the detector plane, obtaining the projected track. The last step involves checking if the track would be bright enough to be detected. For this, tracks for which the mean distance between $\mathbf{r}_{1,2}$ and the camera is larger than a limit distance $\Delta$ are discarded. This limit distance is calculated from the equation \citep{Agarwal2016}

\begin{equation}
    \Delta = \sqrt{\frac{{r_d}^2 \,p\, \Phi(\alpha)\, I_{\odot}}{J\, {r_h}^2}},
\end{equation}

\noindent where $p$ and $\Phi(\alpha)$ are the geometric albedo and phase function of the agglomerate respectively, $I_{\odot}$ the solar flux in the corresponding filter with units of  W m\textsuperscript{-2} nm\textsuperscript{-1}, $r_h$ the heliocentric distance in au and $J$ is the image background brightness, estimated as its median, with units of W m\textsuperscript{-2} nm\textsuperscript{-1}.

\section{Analysis of the detected tracks}\label{sec:realtracks}

\subsection{Orientation}\label{sec:measured_orientation}
\begin{figure}
    \centering
    \includegraphics[width = .9\linewidth]{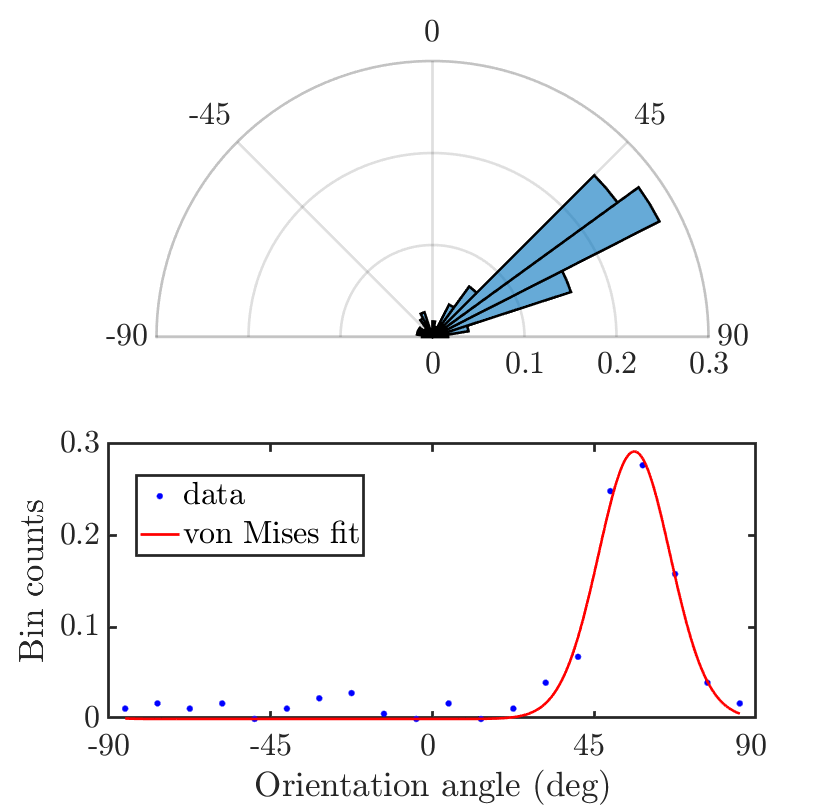}
    \caption{Normalized histogram for the orientation angle of the tracks detected in the image taken in STP092 at a phase angle of 50\degr with the Red filter. The bottom panel shows the von Mises fit obtained for the data. }
    \label{fig:vMfit}
\end{figure}

In the first place, the distribution of the orientation angle of the tracks is analyzed by generating their histograms when grouping them into 20 bins spanning the $[-90\degr:+89\degr]$ interval, and normalizing the histogram such that the sum of the bar heights equals one. Then, a modified von Mises distribution is used to fit the histogram. The von Mises distribution is an approximation of the normal distribution for a periodic domain, expressed by the equation

\begin{equation}
    f(x) = \frac{\exp{\left[\frac{\cos{(x-\mu)}}{\sigma^2}\right]}}{2\pi I_0(1/\sigma^2)},
\end{equation}

\noindent where $\mu$ and $\sigma$ represent the mean and standard deviation respectively, and $I_0$ is the modified Bessel function of the zeroth order. This function is defined over the $[0,2\pi]$ domain, so it is modified in order to match the one of the orientation angles. An example of the normalized histogram for the orientation angles and the von Mises fit can be seen in Fig. \ref{fig:vMfit}.

\begin{figure}
    \centering
    \includegraphics[width = .9\linewidth]{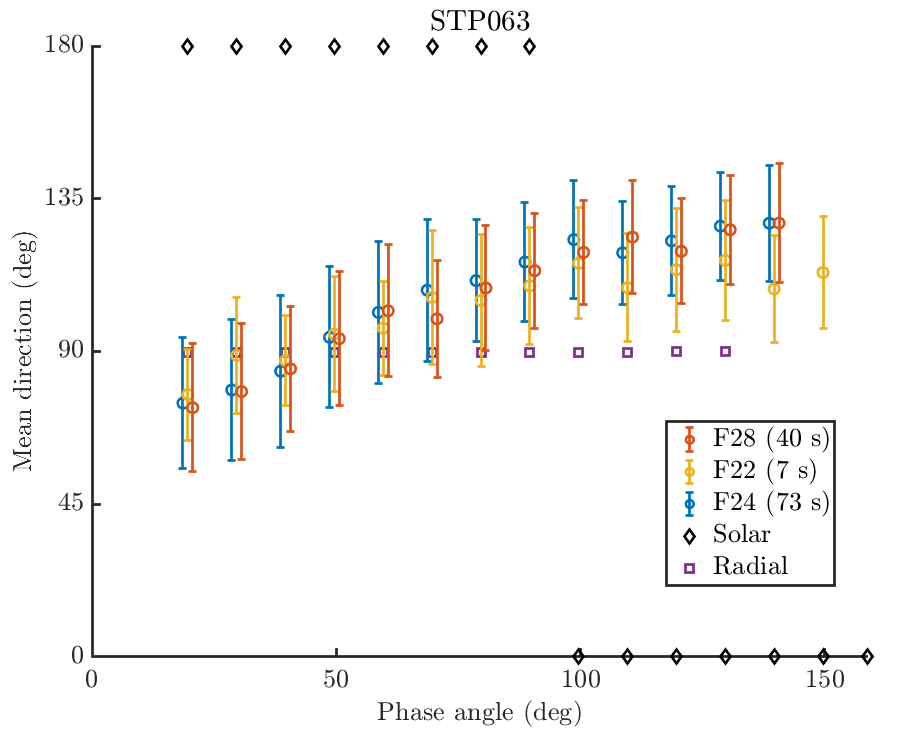}
    \includegraphics[width = .9\linewidth]{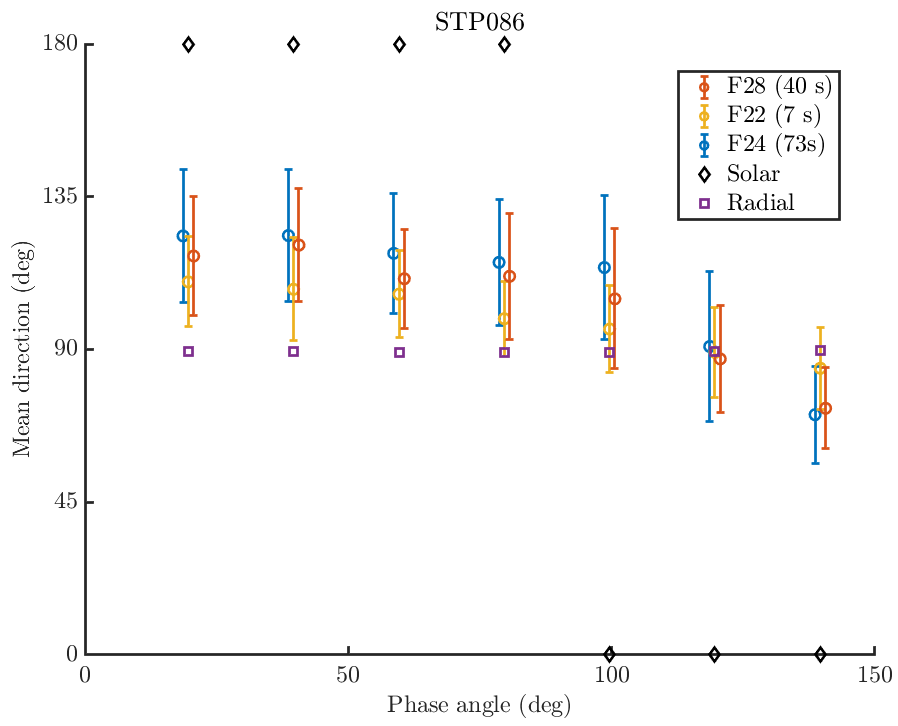}
    \includegraphics[width = .9\linewidth]{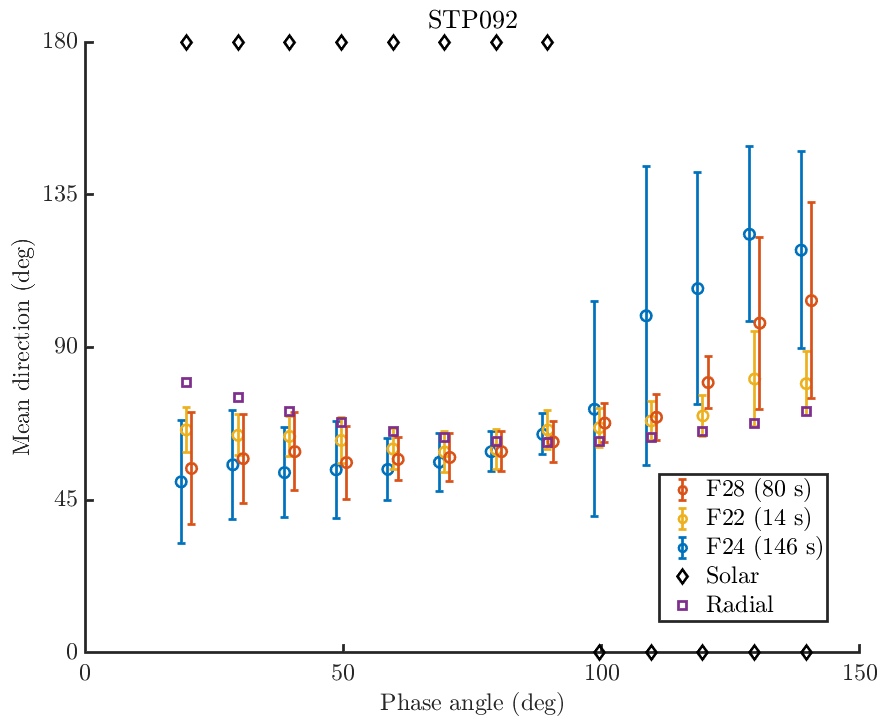}
    \caption{Mean direction of the tracks in the images. The symbol represents the mean and the errorbar the standard deviation of the von Mises distribution fitted to the orientation angle histogram.}
    \label{fig:meandir}
\end{figure}

The mean orientation angle of the tracks in the images as a function of the phase angle can be seen in Fig. \ref{fig:meandir}. We find that there is a difference between the mean direction of the tracks and the radial direction. The radial direction is defined as the direction of the spacecraft--nucleus vector projected into the image plane. From this Figure we notice that this discrepancy shows two particular features. First, the deviation is not constant along each set, but depends on the phase angle. Second, in almost all cases, the deviation from the radial direction depends on the exposure time of the images: the shorter the exposure time, the closer the mean direction of the tracks is to the radial one. In Section \ref{sec:orient_dis} we discuss about the origins of these features.

\subsection{Phase function}

\begin{figure}
    \centering
    \includegraphics[width = \linewidth]{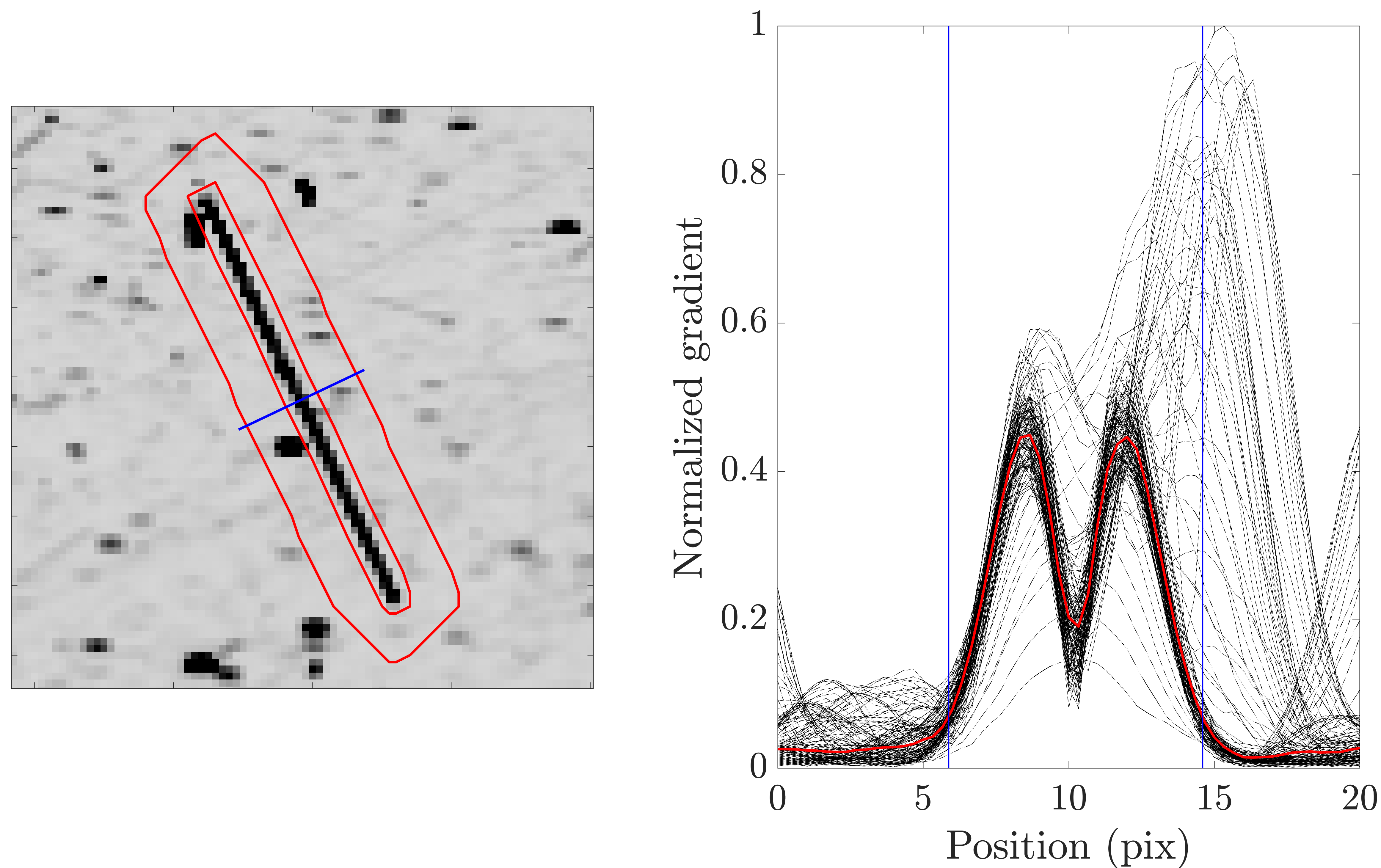}
    \caption{Example of the photometry performed on the tracks. The two stadium shapes on the left panel are obtained by dilating the detected track with a disc. The radius of the disc is obtained from the gradient of the image taken from segments perpendicular to the tracks (right). The blue line in the left panel represent one of the perpendicular segments along which the image gradient is obtained. The black dotted lines show the gradient profile along all the segments, while the median of all profiles is shown with the red solid line. Using these profiles, the total width of the track (blue solid lines) can be found, and it is used as the width of the inner aperture. The outer aperture has a fixed total width of 20 pixels.}
    \label{fig:photometry}
\end{figure}

For computing the phase function of the tracks, the photometry of all tracks completely enclosed in the image was performed. The method is similar to the one described in \citet{Guttler2017}, namely performing a morphological dilation of the original track with two ring sizes, in order to obtain two stadium shapes enclosing the track. The size of the discs used for the dilation are estimated from the local gradient image $G$ (Fig. \ref{fig:photometry}). The gradient image $G$ is calculated as $G = \sqrt{G_x^2+G_y^2}$, where $G_{(x,y)}$ are the directional gradients obtained using a Gaussian kernel. The pixels contained in the inner shape are summed to obtain the total track brightness, while the background is estimated as the median value of the pixels between both shapes, and subtracted from the brightness of the central shape.

\begin{figure}
    \centering
    \includegraphics[width = \linewidth]{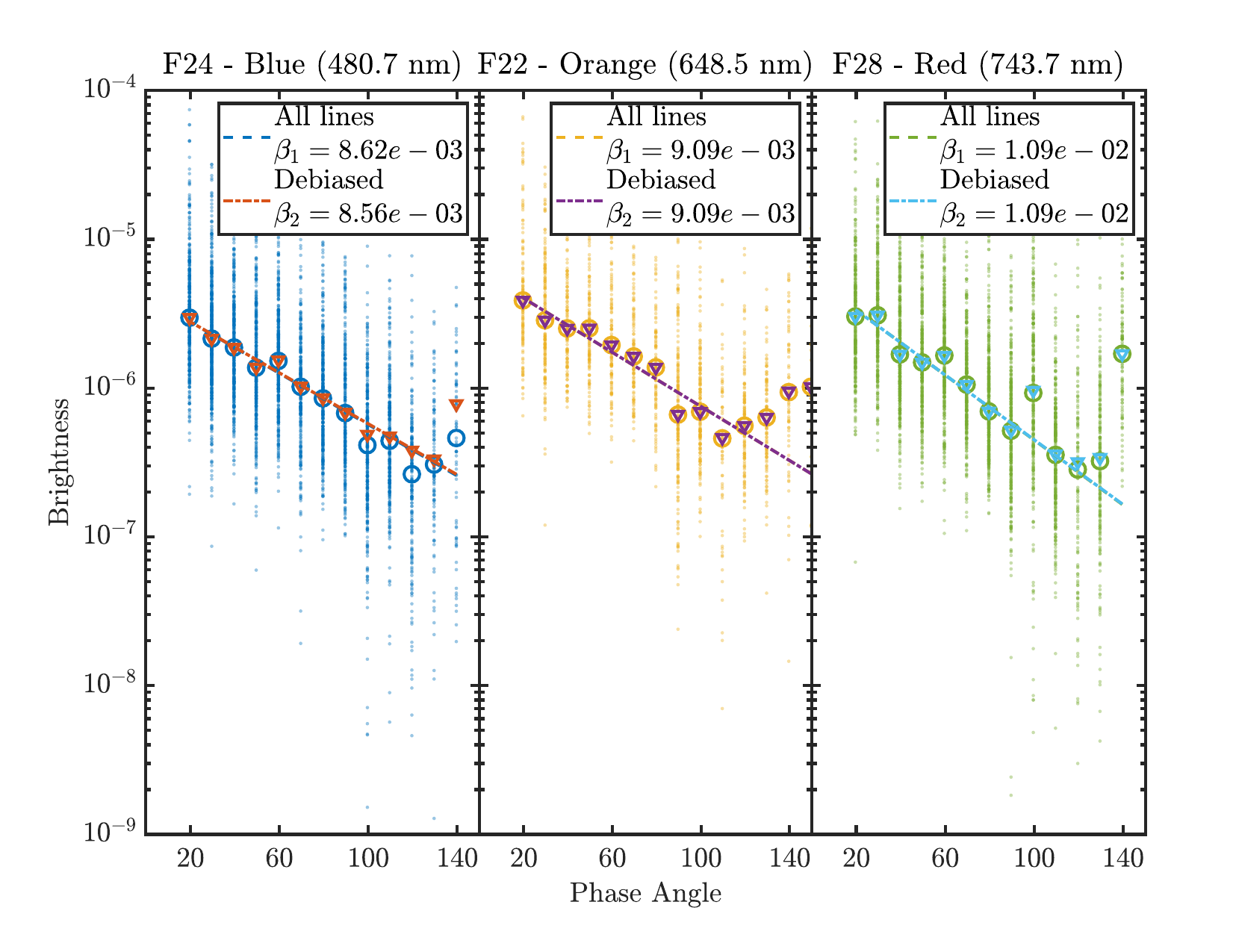}
    \includegraphics[width = \linewidth]{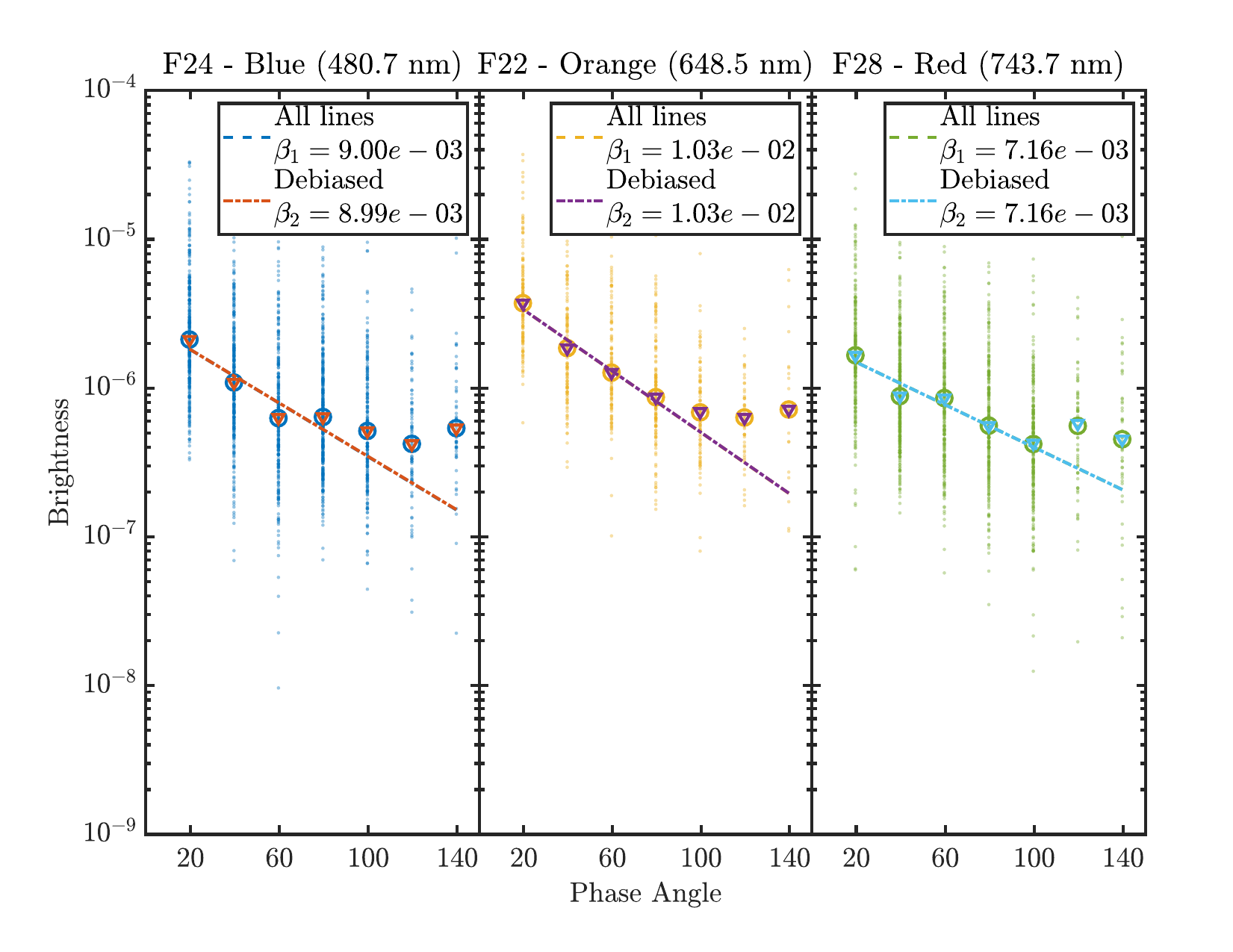}
    \includegraphics[width = \linewidth]{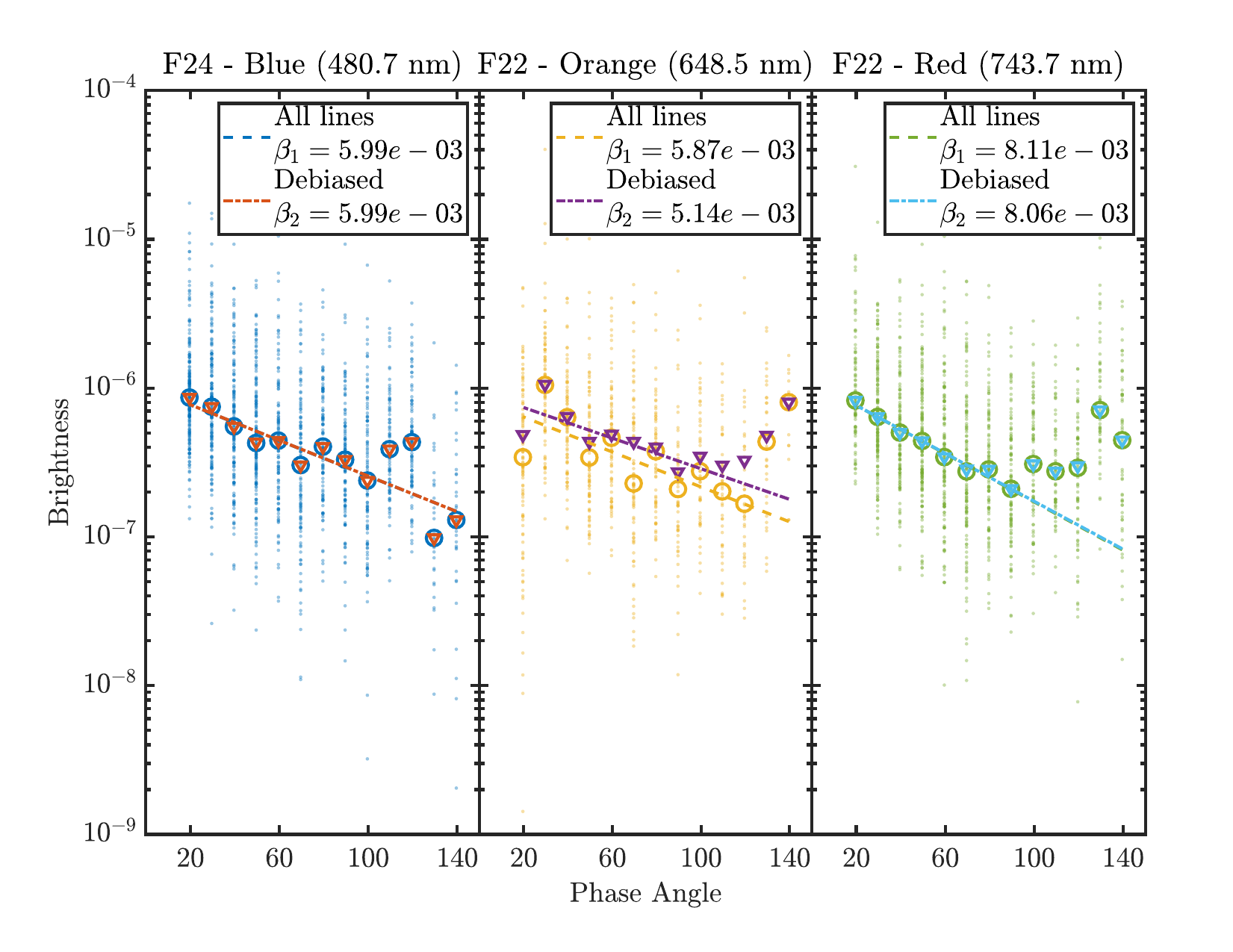}
    \caption{Phase function of the observed agglomerates. From top to bottom are the results for STP063, STP086 and STP092. The light colored dots indicate the integrated reflectance for each track, while the colored triangles show the median for each phase angle. The colored circles represent the median of the integrated reflectance after filtering out dim tracks.}
    \label{fig:phasefun}
\end{figure}

The brightness of all the detected tracks entirely contained in the images are represented by the small, coloured dots in Fig. \ref{fig:phasefun}. This quantity does not depend on the apparent speed of the agglomerates (except in the case the apparent speed is small and the track is shorter than 10 pixels, see Sec. \ref{sec:detMethod}), but only on their distance to the camera, size and scattering properties. The image acquisition method consists of keeping the spacecraft in the same position and rotate it to obtain images at different phase angles. We assume that all observed agglomerates are in the vicinity of the spacecraft and that the population of dust agglomerates generating the tracks for a given set is similar for all phase angles. From this we can estimate the phase function of the agglomerates performing a statistical analysis of the brightness, by assuming that the most representative value for a certain phase angle is the median of all tracks for that image. 

However, the set of track brightness is twofold biased. First, as mentioned before images taken at high phase angles are contaminated by straylight, which means that the background brightness is much higher than in the images at low phase angle. For this reason, faint tracks cannot be detected in high phase angle images as they blend into the background, so the sample is biased towards brighter tracks, as can be seen in Fig. \ref{fig:phasefun}. For the rest of this phase function analysis, tracks obtained from images taken at phase angles greater than 120\degr will be discarded. Secondly, the scattering phase function values are higher for low phase angles, so fainter agglomerates can be detected in them. This effect introduces a bias to the phase function derived from the detections. For overcoming this issue, we will adopt an iterative process. Following the results of \citet{Fornasier2015} and \citet{Guttler2017}, we fit the median values of the track brightness using an exponential function of the form $R(\alpha) = A\times\exp(-\beta\times\alpha)$. Using this result as a preliminary phase function, we look for the faintest track in the images taken at the highest phase angle, and extrapolate its brightness for the remaining images. This extrapolated value is used as a lower threshold for the brightness of the tracks considered for the second iteration step, discarding all fainter tracks. The fitting is then repeated and the coefficients are calculated again for the debiased sample.

The chosen expression for the phase function provides a good fit of the values after removing the values for phase angles greater than 120\degr from the sample: the mean value of the coefficient of determination $R^2$ for all samples is $0.83$. This confirms the results of \citet{Fulle2018}, who show the characteristic U-shape function found for the dusty coma phase function \citep{Bertini2017} is valid for particles with radius $r<1.25$ mm, smaller than the ones observed in the OSIRIS data used in Section \ref{sec:observations}. However, we find that the mean value for all the sets is $\beta=8.2\times 10^{-3}$, around five times smaller than that found by previous works focusing on the comet nucleus. Following the classic theory by \citet{Lumme1981I,Lumme1981II}, this can be explained by shadowing due to different surface roughness. Even when the nucleus phase function is corrected for the self-shadowing, the total brightness depends on the pixel resolution, since non-resolved shadows cannot be corrected and affect the final result (see fig. 1 in \citealt{Hasselmann2021}). Because of this, a lower $\beta$ value for the coma agglomerates is consistent with the smaller size of the dust agglomerates in the coma compared to that of the nucleus. 

While analyzing the time evolution of $\beta$, we find that the mean value found for all three filters is $9.5\times 10^{-3}$, $8.8\times 10^{-3}$ and $6.4\times 10^{-3}$ for the sets STP063, STP086 and STP092 respectively. Using the same argument as before, this can be explained as a reduction of the median size of the agglomerates in the coma, which is consistent with the increase of the heliocentric distance of the comet. 

\section{Comparison with the model and discussion}\label{sec:comparison}

\subsection{Orientation angle distribution}\label{sec:orient_dis}

\begin{figure}
    \centering
    \includegraphics[width=\linewidth]{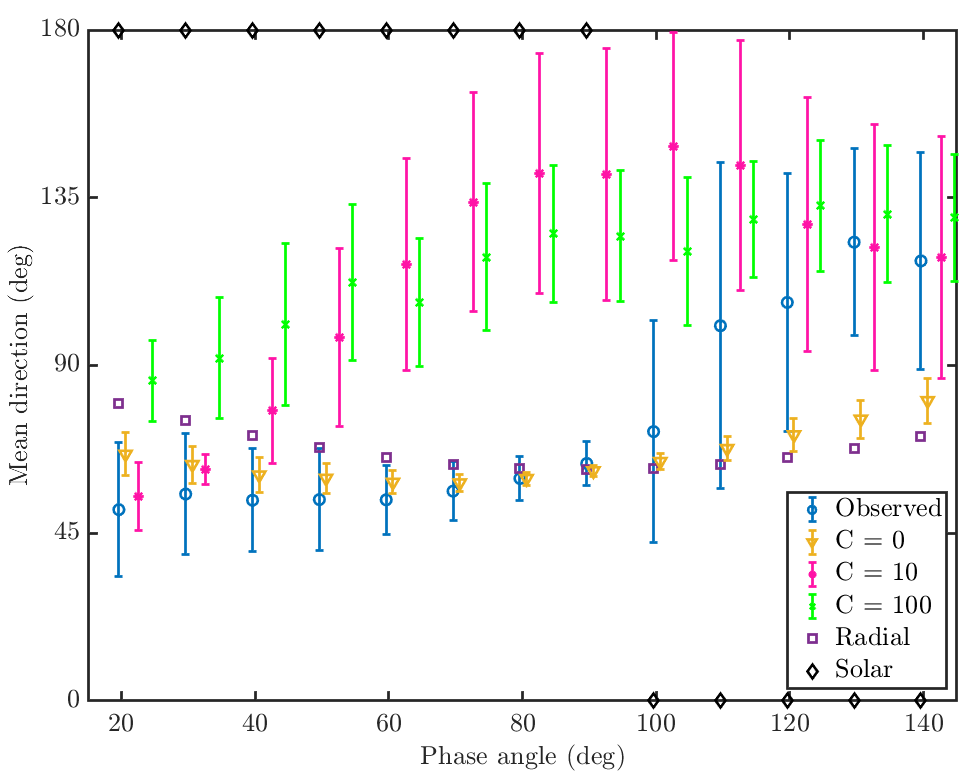}
    \caption{Mean direction of the tracks in the F24 images for STP092, compared against the mean directions for the simulated trajectories with radiation pressure factors of $C=$ 0, 10 and 100. Results for $C=1$ were almost identical to the ones for $C=0$, so were not included in this plot.}
    \label{fig:projef}
\end{figure}

In order to test if the measured directions introduced in Section \ref{sec:measured_orientation} can be explained by a projection effect, we create synthetic images corresponding to the STP092 observation geometry, but without taking into account the gaseous drag. Since the results depend on density and size mainly through the gas drag, the choice of these parameters does not affect the results excessively. For this test case, we use agglomerates with $\rho = 100$ kg m\textsuperscript{-3}, $r_d = 1$ cm which are ejected from the nucleus with a most probable speed of $1$ m s\textsuperscript{-1}. In order to check the relevance of the radiation pressure on this effect, we compute the trajectories of agglomerates in initially radial trajectories under the effect of the gravitational and radiation pressure forces, but including a multiplicative factor $C$ for the latter. Fig. \ref{fig:projef} displays the same plot as in Fig. \ref{fig:meandir} for the aforementioned set, but with the mean directions for various $C$ values superimposed. In the purely gravitational case ($C=0$), the agglomerates move in radial trajectories, but even so, the deviation can be observed. This effect can be explained by a simple projection effect: here, the radial direction is defined as the projection onto the image of a vector joining the nucleus and the spacecraft. Since the agglomerates are not in the same position as the spacecraft, the projection of their own (local) radial directions is not necessarily parallel to the one at the spacecraft.

Although the projection effect can explain the trend of the deviation from radial direction as a function of the phase angle for the purely gravitational ($C=0$) case, this effect alone is not sufficient to explain the absolute value of the deviation at phase angles larger than 120\degr. Additionally, the observed angular dispersion is larger than the one found with this model for all phase angles. However, a value of $C\gg1$ can account for the mean direction as well as the angular dispersion in the high phase angle region.

From a physical perspective, several processes can be invoked in order to explain an enhancement of the radiation pressure. For example, an agglomerate mass versus cross section ratio smaller than the one used for the integration, either caused by a lower density or a nonspherical shape, can explain the higher radiation pressure effect. Also, additional forces parallel to the solar direction, such as outgassing from slowly rotating agglomerates \citep{Kelley2013}, can account for the effect. However, it is worth noticing the limitations given by the choice of boundary conditions. In order to reduce the time required for the simulations, the integration domain limit is set to an altitude 20 km higher than that of the spacecraft. In the case that some agglomerates that are ejected from the nucleus decelerate and fall back at altitudes above the domain limit, their trajectories would not be taken into account by our model. The inclusion of these agglomerates may be able to modify the results, even for typical values of radiation pressure forces. 

This finding represents a nuance with respect to previous results. \citet{DellaCorte2016} and \citet{Longobardo2019,Longobardo2020} report radial trajectories for particles analyzed by GIADA, but the smaller size of these particles compared to those that OSIRIS is able to observe (see Sec. \ref{sec:param_opt}) may explain this feature, since they are more affected by the gaseous drag. In addition, \citet{Longobardo2020} proposes that the motion of the particles could only be considered to be radial up to altitudes of $\sim40$ km, since the radiation pressure plays an important role for higher altitudes. Likewise, \citet{Gerig2018} find that dust agglomerates observed by OSIRIS follow a free-radial outflow from the nucleus for distances larger than $12$ km from it, but their analysis is limited to altitudes up to $40$ km. On the other hand, \citet{Frattin2021} analyze similar OSIRIS images, with tracks generated by the motion of dust agglomerates. As in our case, they find that most of the agglomerates have trajectories close to the radial direction, and interpret the remaining ones as a population of objects on bound orbits around the nucleus. 

As a summary, the explanation for the track direction presented in this work proposes that agglomerate trajectories have a clear general orientation. Like the interpretation proposed in previous works, we find that this general orientation is close to the radial direction once the projection effect is taken into account. However, for phase angles greater than 120\degr, both the most probable orientation angle and its dispersion cannot be well reproduced by radial trajectories only. For explaining these trajectories, forces other than the gravity of the nucleus (e.g. radiation pressure) or, following the explanation by \citet{Frattin2021}, an increased proportion of agglomerates in bounded orbits, must be considered. 

\subsection{Dust parameter optimization}\label{sec:param_opt}

As was explained in Section \ref{sec:synthIm}, the generation of synthetic images is done based on the trajectories computed for individual dust parameters combinations. However, no single parameter combination will fully represent the observations, since the analyzed OSIRIS images contain tracks generated by a variety of different agglomerates. For overcoming this issue, we will assume that the distribution of track properties obtained from the real images can be expressed as a linear combination of those in the synthetic ones. This implies that interactions between different types of agglomerates, either directly through collisions or indirectly mediated by the gas in the coma, are considered not relevant for their dynamical evolution.

We will focus on comparing the distribution of track properties in the length--orientation angle space. In order to compare the distributions of these properties with those obtained from the real images, we generate normalized histograms with equal bin ranges for tracks detected in both types of images. Mathematically, this is equivalent to creating a 2D matrix in which each element represents the number of tracks with a specific combination of length and orientation angle, and normalizing it by the total number of tracks in the image. Then, for each observation geometry there exist two types of histograms, the one obtained from the real image $\mathbf{x}$, and from the synthetic ones $\mathbf{y}_i$, where $i$ represents the different dust properties used for the simulations. We generate the \textit{master} synthetic histogram $\mathbf{Z}$ from a linear combination $\mathbf{Z} = \sum_i K_i \mathbf{y}_i$. The coefficients $K_i$ for the linear combination, which roughly represent the preponderance of agglomerates with certain properties in the image, are chosen in such a way that they minimize the $\chi^2$ distance between the real $\mathbf{x}$ and master synthetic $\mathbf{Z}$ histograms. The $\chi^2$ distance measures the distance between two histograms with $N$ bins, and is defined as

\begin{equation}
    \chi^2 = \frac{1}{2}\sum_{j=1}^{N} \frac{(x_j - Z_j)^2} {(x_j + Z_j)},
\end{equation}

\noindent where $x_j$ and $Z_j$ represent the value of each bin for the real $\mathbf{x}$ and master synthetic $\mathbf{Z}$ histograms respectively.

\begin{figure}
    \centering
    \includegraphics[width = \linewidth]{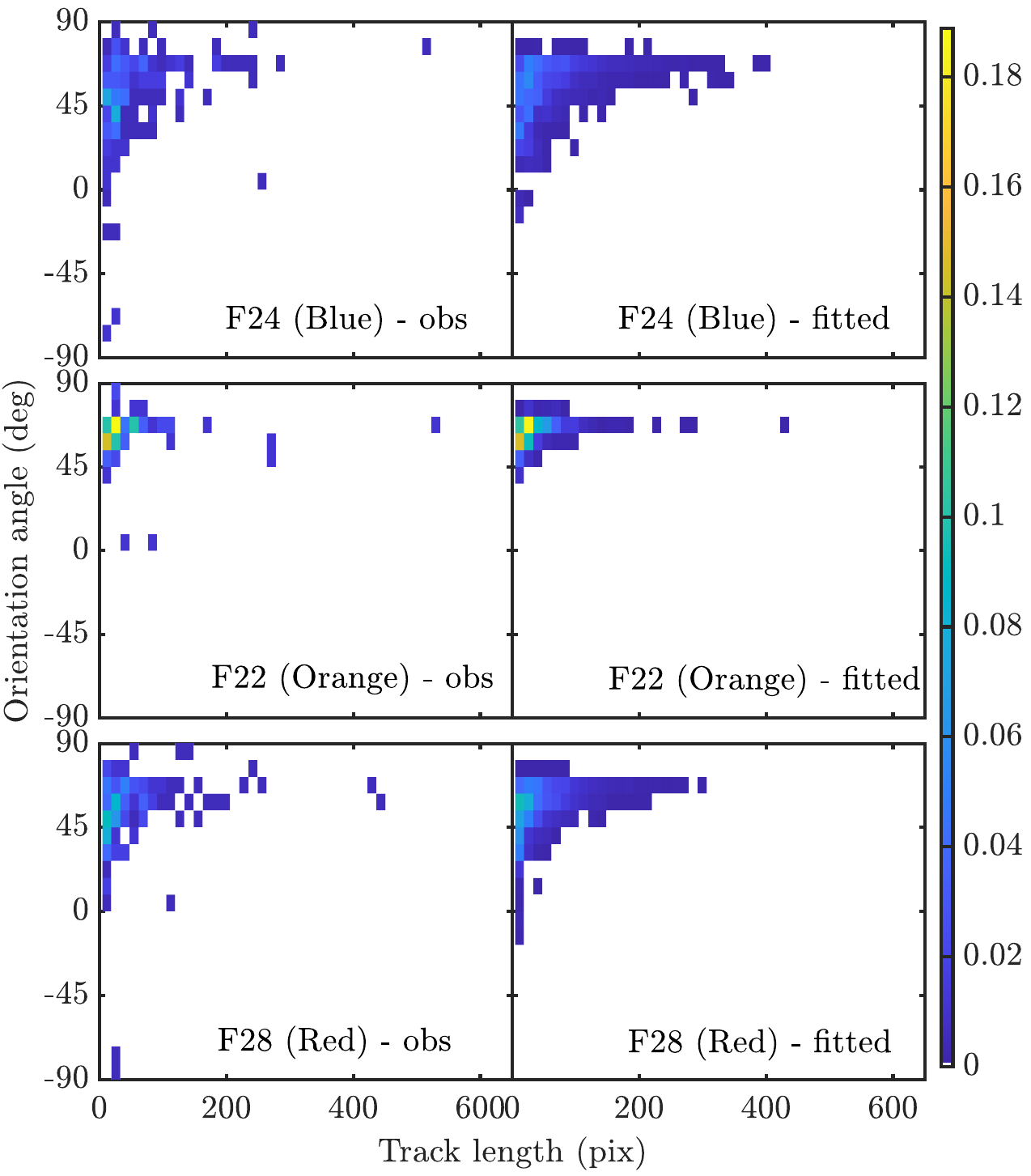}
    \caption{Observed and fitted distribution of the tracks in the orientation angle versus track length found for the set taken at STP092 with a phase angle of $\alpha=$30\degr. The rows represent the filter used for the image acquisition (from top to bottom Blue, Orange and Red), while the left and right columns show observed and fitted distributions respectively. The color code represent the number of tracks found in that particular bin, normalized by the total number of tracks in the image.}
    \label{fig:fithist}
\end{figure}

The $K_i$ coefficients giving the best-fitting are found using an iterative linear least-squares solver with random initial values, while the constraints for the coefficients are $K_i>0$ and $\sum_i K_i = 1$. For simplicity, we computed the $\chi^2$ distance for all the individual histograms $\mathbf{y}_i$ and calculated the master synthetic histogram $\mathbf{Z}$ using only a subset of 100 synthetic images with the best results. In order to check whether the image subset choice affects the results, we repeated the fit with different number of synthetic images. We found that performing the fit with this 100 synthetic images provided the best compromise between accuracy of the results, showed by the residual after the fit, and the execution time. An example of the best fit histograms obtained using this method can be seen in Fig. \ref{fig:fithist}. 

\begin{figure}
    \centering
    \includegraphics[width=.9\linewidth]{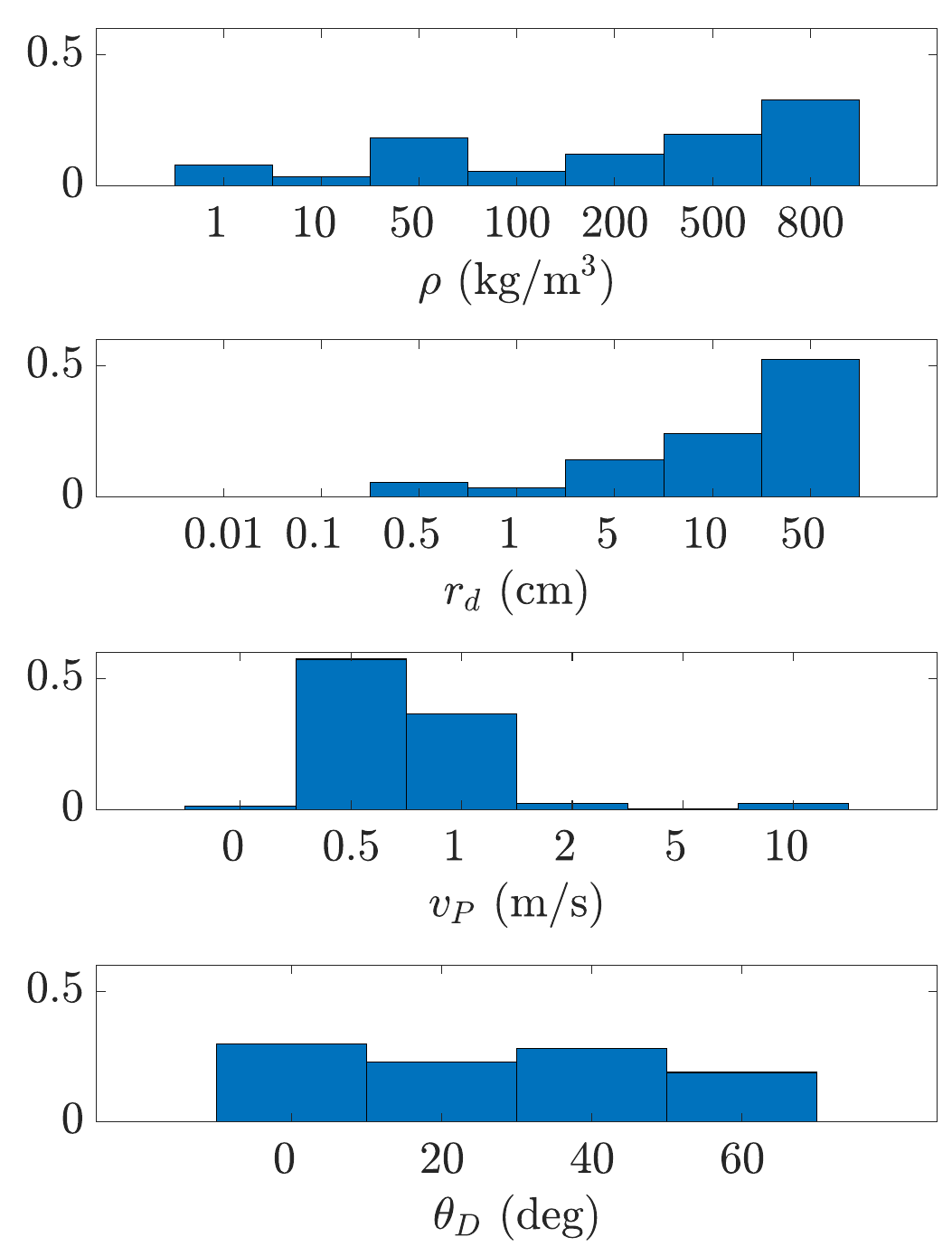}
    \caption{Weighted distribution of the parameters found for the image taken at $\alpha=120\degr$ with the Blue filter in the set STP092.}
    \label{fig:ind_hist}
\end{figure}

We use these $K_i$ values to obtain a weighted distribution of the parameters used in the simulations. Fig. \ref{fig:ind_hist} shows an example of this weighted distribution for an image from the set STP092 taken at a phase angle $\alpha=120\degr$ with the Blue filter. This distribution is obtained by grouping the synthetic images by their values of a certain parameter (in the case of Fig. \ref{fig:ind_hist}, density, agglomerate radius, initial speed and initial direction), and summing the $K_i$ values of the groups. We found that varying the number of synthetic images used for the fit does not change significantly these results.

\begin{figure*}
    \centering
    \includegraphics[width = .9\linewidth]{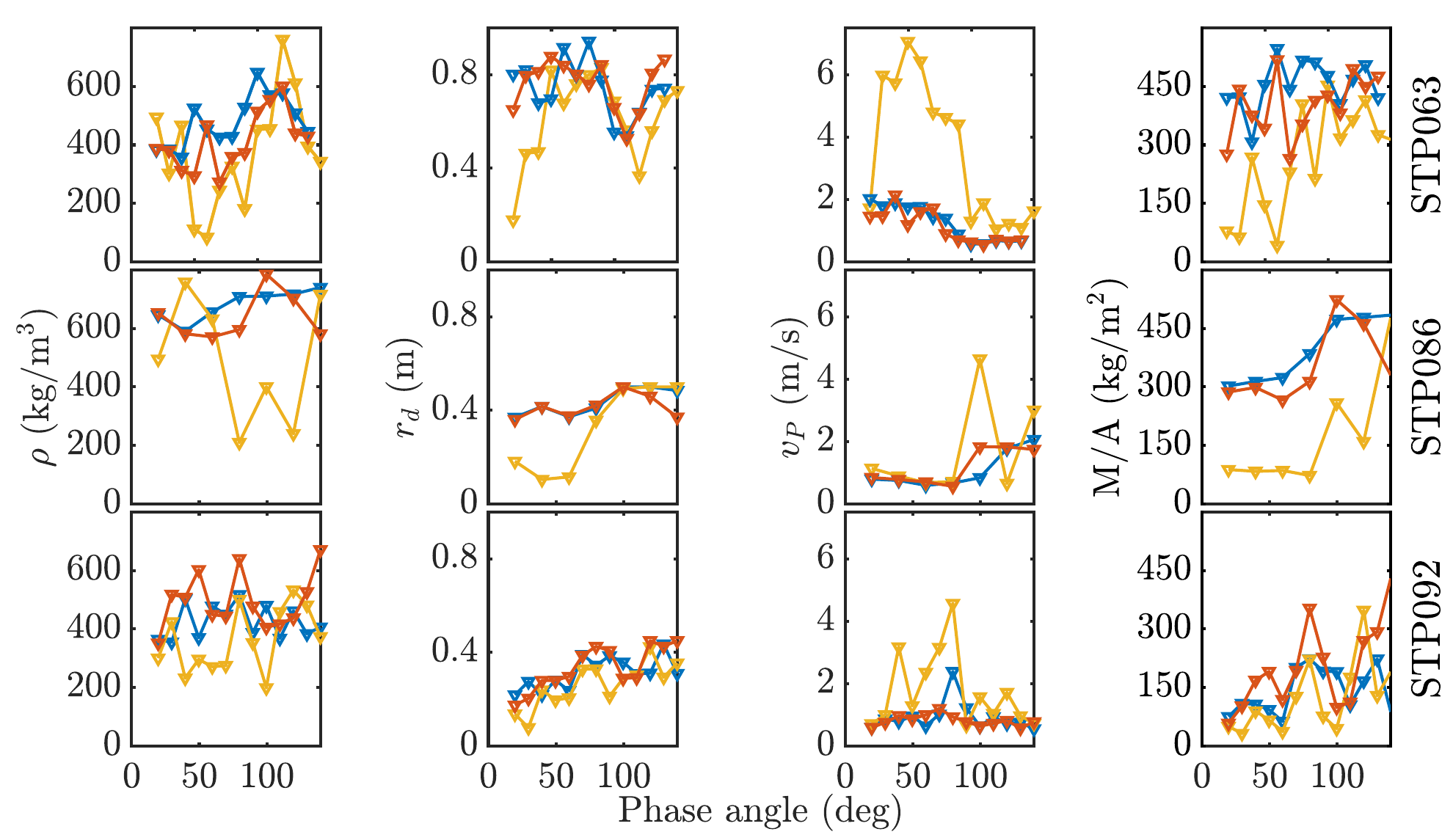}
    \caption{Weighted mean for the dust parameters. The four selected parameters, density, radius, initial speed and $M/A$ ratio are displayed in columns from left to right, while the results for STP063, STP086 and STP092 are shown in the top, middle and bottom rows respectively. Line colors indicate the used filter as in Fig. \ref{fig:meandir}.}
    \label{fig:meanpar_ev}
\end{figure*}

The weighted means can be found from the mentioned distributions. Fig. \ref{fig:meanpar_ev} shows the dependence of these means with the phase angle at which the images were taken. We plot 4 different parameters: agglomerate density, radius, most probable initial speed and mass over area ratio. The mass over area ratio $M/A=(4/3\pi\rho {r_d}^3)/(\pi{r_d}^2)=4/3 \rho r_d$ is useful for quantifying the effect of the radiation pressure and gaseous drag over the agglomerate dynamics, since both the $F_G/F_R$ and $F_G/F_D$ ratios depend linearly on it. This model is not able to provide tight constraints for the density, and hence neither for the $M/A$ ratio, but a clear trend can be seen for the remaining parameters. Lastly, as can be seen in the bottom panel in Figure \ref{fig:ind_hist}, the results are independent from the choice of the angle between initial velocity and local normal $\theta_d$, so are not shown here. This is because while the initial velocity of the dust agglomerates is not radial, the initial velocity of the gas is, so the gaseous drag force, which is very strong due to the high gas density in the vicinity of the nucleus, cancels out the possible effect of this non-radial initial velocity.

As can be seen in Figure \ref{fig:meanpar_ev}, the results for the images obtained using the Blue and Red filters match with each other, while the ones for the Orange filter show a larger deviation. As mentioned in Section \ref{sec:detMethod}, the exposure times used for this type of images introduces a bias towards agglomerates that are faster/closer to the camera, and may be responsible for the observed discrepancies.

\begin{figure}
    \centering
    \includegraphics[width = \linewidth]{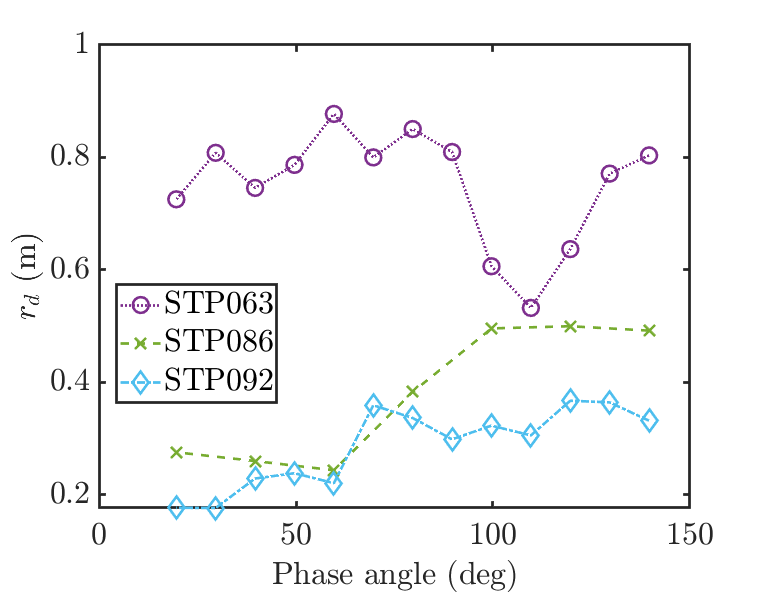}
    \caption{Weighted mean chunk size obtained from the combinations of Blue and Red filters as a function of the phase angle. It can be seen that the chunk size increases with smaller heliocentric distances.}
    \label{fig:combined_rd}
\end{figure}

It can be noted that the mean sizes for the chunks follow the expected trend: the mean chunk sizes are larger for the sets taken closer to the perihelion, and they are larger for increasing phase angles, that is, looking into the dayside of the coma. In order to increase the sample size, the combined results for the chunk size as a function of phase angle obtained from the fits for the Blue and Red filters are shown in Figure \ref{fig:combined_rd}. However, the sizes are much larger than the theoretical maximum liftable size found using the model by \citet{Fulle2020}. In contrast, \citet{Gundlach2020} and \citet{Ciarniello2022} show that CO\textsubscript{2} ice sublimation is the main driver of the activity of chunks with sizes $\gtrsim$ 10 cm. This is because the water sublimation front is located at shallower depths from the surface, so it can only build up enough pressure to overcome the material internal strength and eject chunks at these shallow depths. On the other hand, the CO\textsubscript{2} sublimation front is located deeper, allowing to eject larger chunks. It is important to notice that the studied case assumes that the agglomerates are ejected when the gas pressure overcomes the material tensile strength. Therefore this model is not able to analyze the case where a detached agglomerate resting on top of the surface manages to gain an initial impulse and gets lifted. 

The scenario where the chunks are ejected via CO\textsubscript{2} sublimation is consistent with the observed nonzero mean initial speed of the agglomerates found in our simulations. Since CO\textsubscript{2} sublimation is not included in our model, this can be represented as an initial kick to the chunks, making it possible for them to be lifted. Once in the coma, the agglomerates evolve dynamically under the influence of the mentioned forces, in particular gas drag. However, since the CO\textsubscript{2} production rate is around one order of magnitude lower than that of H\textsubscript{2}O, the latter controls the drag force, and the assumption of a coma composed by water vapour is still valid.

Regarding the initial speed values, we observe that initial velocities as derived from histogram fitting are larger for the set from STP063, that is, the closest one to perihelion. Moreover, in the purely gravitational case (i.e. without radiation pressure nor gas drag), the initial speed required for reaching the altitudes at which the spacecraft was located in all the three analyzed sets is $\simeq 0.80$ m s\textsuperscript{-1}. Also for all three data sets, the initial velocities found via the histogram fitting (Fig. \ref{fig:meanpar_ev}) are sufficient for the chunks to reach the spacecraft altitude.

Hence, the mere presence of these agglomerates in the FOV does not imply that they have been significantly accelerated by gas drag after leaving the nucleus surface. Due to their large size, it is unclear whether gas drag has any relevance to the dynamic evolution of these chunks.

We use an indirect approach to estimate the effect of gas drag on the observed chunks, based on the fraction of them having bound orbits and the distribution of initial velocities of agglomerates inside the FOV. If we assume the gas drag does not influence the dynamics of the chunks, the initial speed needed to reach the FOV (0.80 m s\textsuperscript{-1}) is close to the escape speed for the spherical nucleus (0.82 m s\textsuperscript{-1}). Since the initial velocities are taken from the Maxwell--Boltzmann distribution given in equation \ref{eq:maxw}, we calculate the probability that an agglomerate has an initial speed sufficient for reaching the FOV, but still smaller than the escape speed. We find that this condition is fulfilled by only a small proportion (<1\%) of agglomerates in this purely gravitational case. 

However, if we analyze the energy of the agglomerates that intersect the FOV in our simulations (which include the effect of gas drag), we find a much higher proportion of bound orbits. This number is highly variable between data sets, ranging from 0 to 30 per cent, with a mean of 12. We interpret this finding such that the majority of the bound chunks were initially lifted with speeds too small to reach the FOV, but were subsequently accelerated towards crossing the FOV by gas drag. 

This interpretation is supported by 32 per cent of the chunks that reach the FOV having initial speeds lower than needed to reach the spacecraft altitude. All this implies that the dynamics of the large chunks found in our simulations is still affected by gas drag.

\begin{figure}
    \centering
    \includegraphics[width = .9\linewidth]{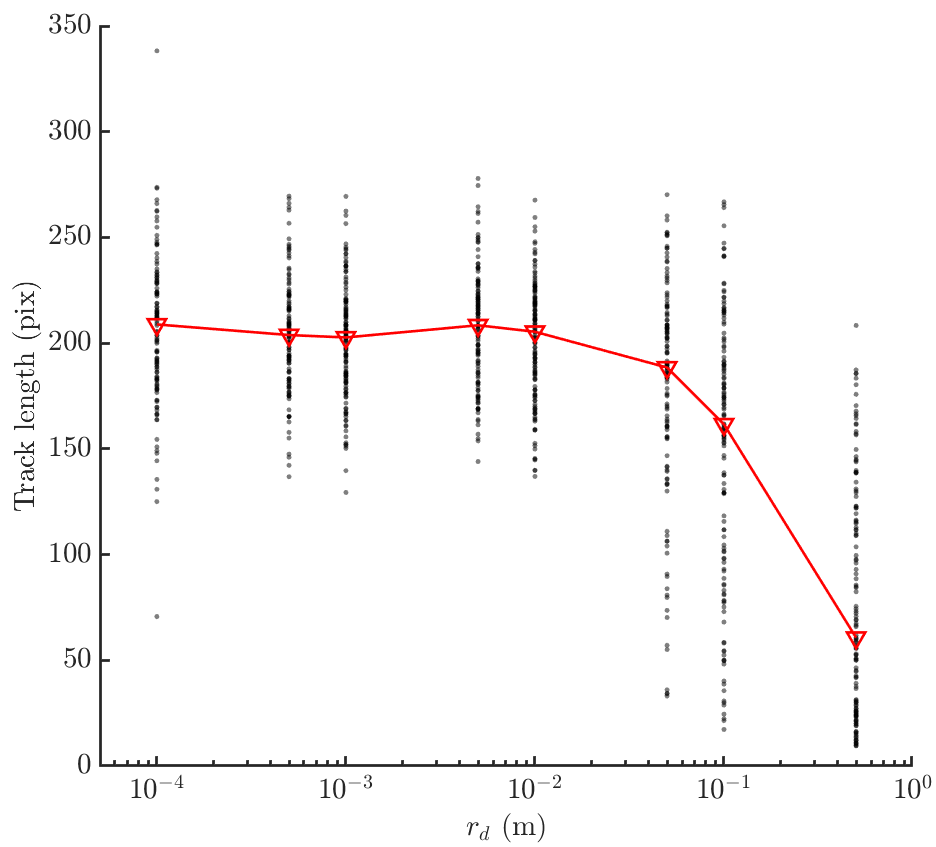}
    \caption{Median track length found in synthetic images for different values of particle radius. Each black dot represents the length median value of all the track found in a synthetic image. Red symbols show the median value of all images grouped by the particle size used for the simulation. A clear trend can be seen where larger agglomerates generate shorter tracks.}
    \label{fig:medianlen}
\end{figure}

It is important to note that the chunk sizes found in this work are larger than the ones found by \citet{Frattin2021} for the same type of images. Agglomerate sizes compatible with \citet{Frattin2021} would have too high velocities in our model to be compatible with the observed length of tracks. Fig. \ref{fig:medianlen} shows the median track lengths found in the synthetic images as a function of the agglomerate size, where it can be seen that for small particles, the tracks are longer. The reason for this behaviour is that after acceleration by gas drag, the ratio $F_G/F_D \propto r_D$ for a fixed particle density. This implies that smaller particles are more susceptible to the action of the gas drag, and can acquire higher velocities generating longer tracks. For agglomerate sizes compatible with \citet{Frattin2021}, the tracks in the synthetic images are longer than those in our OSIRIS images, so our data can only be reproduced by larger chunks. This effect is particularly noticeable for the case of STP063, where the gas flow was so intense that additional simulations with larger chunks had to be carried out (see Table \ref{tab:dustprop}). 

However, there may exist other effects that slow down the agglomerates. For example, the effect of solar gravity that was not taken into account in the dynamical simulations could provide an alternative way of producing shorter tracks. Since the simulations were carried out in the non inertial nucleocentric reference frame, the net force acting over the agglomerates in this frame is the difference between solar gravity force over the agglomerate and the nucleus, i.e. the tidal force. This force increases linearly with the nucleocentric distance, and due to the observation geometry (the nucleus--spacecraft vector forming $\simeq$90\degr with the radial direction), its direction at the position of the spacecraft is radial pointing to the nucleus, effectively increasing the value of the nucleus gravity acceleration and slowing even further the observed dust agglomerates. Assuming that all the agglomerates present in the images are at the same height from the nucleus as the spacecraft, and using the approximated expression for the tidal acceleration $a_T = \mathcal{G} M_{\odot} r / r_h^3$, the ratio between tidal and nucleus gravity forces $F_T/F_G$ is equal to 0.19, 0.02 and 5.7$\times 10^{-3}$ for the sets STP063, STP086 and STP092 respectively. Then, the tidal effects may play a relevant role in the dynamical evolution of the dust agglomerates, mainly for the first set . 

\section{Conclusions}\label{sec:conclusion}

We developed a semi automatic method to detect tracks generated by agglomerates moving in front of the OSIRIS camera onboard Rosetta. This method exploits the fact that the agglomerates move in front of the camera generating the particular track pattern. We applied this method to three different image sets taken with the NAC camera composed by a total of 105 images, and detected 20033 tracks. 

We analyzed the photometric data obtained from those tracks, and found that the agglomerates' phase functions do not show the characteristic U-shape found for the phase function of the coma \citep{Bertini2017}, but rather follow the same exponential trend as the one shown by the nucleus \citep{Fornasier2015,Guttler2017}. Following \citet{Fulle2018}, this establishes a lower limit for the agglomerate sizes at $r>1.25$ mm. The value of the phase function exponent $\beta$ found for the agglomerates is smaller than the nucleus one, consistent with the difference in roughness scales between both samples. We also observed that the $\beta$ value decreases for increasing heliocentric distances, indicating a decrease in the median size of the agglomerates detected in the coma. 

We used a simplified dynamical model in order to create synthetic images that reproduce the observations. We solved the inverse problem to find the values characterizing the dust that best reproduce the observed tracks. Using this method we could impose a loose constraint in the density ($\rho$ = 200 -- 800 kg m\textsuperscript{-3}), but tighter ones for the initial velocities ($v_P\simeq 1$ m s\textsuperscript{-1}) and chunk radii (several dm). Both the initial velocities and chunk radii vary for different heliocentric distances, consistent with the gaseous production rate and the observed phase function. 

Even when the radii obtained by the comparison between the observation and the dynamical model only provide an upper limit, the activity model used here cannot provide the required pressure needed to lift agglomerates of such sizes. Instead, it is necessary to invoke other source of gas like CO\textsubscript{2} \citep{Gundlach2020} in order to explain the ejection of those chunks. 

We also showed that other dynamical effects such as solar gravity may play an important role in determining the dynamics of the agglomerates, principally for sets taken closer to perihelion, where the combination of the small heliocentric distance with the high spacecraft altitude makes the agglomerates seen by the spacecraft much more susceptible to its effect. In order to better model the dynamics of the agglomerate, this effect must be taken into account for future works.

\section*{Acknowledgements}
We thank the referee for his constructive suggestions that significantly helped to improve the quality of this manuscript. We thank Nick Atree, Yuna Kwon, Manuela Lippi, Johannes Markannen, Raphael Marschall and Marius Pfeifer for our fruitful discussions. OSIRIS was built by a consortium of the Max-Planck-Institut für Sonnensystemforschung, Göttingen, Germany; the CISAS University of Padova, Italy; the Laboratoire d’Astrophysique de Marseille, France; the Instituto de Astrofísica de Andalucia, CSIC, Granada, Spain; the Research and Scientific Support Department of the European Space Agency Noordwijk, The Netherlands; the Instituto Nacional de Técnica Aeroespacial, Madrid, Spain; the Universidad Politécnica de Madrid, Spain; the Department of Physics and Astronomy of Uppsala University, Sweden; and the Institut für Datentechnik und Kommunikationsnetze der Technischen Universität Braunschweig, Germany. The support of the national funding agencies of Germany (DLR), France (CNES), Italy (ASI), Spain (MEC), Sweden (SNSB), and the ESA Technical Directorate is gratefully acknowledged. We thank the Rosetta Science Ground Segment at ESAC, the Rosetta Missions Operations Centre at ESOC and the Rosetta Project at ESTEC for their outstanding work enabling the science return of the Rosetta Mission. This work used the Scientific Compute Cluster at GWDG, the joint data center of Max Planck Society for the Advancement of Science (MPG) and University of Göttingen. The authors acknowledge funding by the ERC Starting Grant No. 757390 Comet and Asteroid Re-Shaping through Activity (CAstRA). PL conducted the work in this paper in the framework of the International Max-Planck Research School (IMPRS) for Solar System Science at the University of Göttingen. JA acknowledges funding by the Volkswagen Foundation. 

%%%%%%%%%%%%%%%%%%%%%%%%%%%%%%%%%%%%%%%%%%%%%%%%%%
\section*{Data Availability}

The data underlying this article are available at the Planetary Science Archive of the European Space Agency under \url{https://www.cosmos.esa.int/web/psa/rosetta}



\bibliographystyle{mnras}
\bibliography{references} % if your bibtex file is called example.bib

\begin{thebibliography}{}
\makeatletter
\relax
\def\mn@urlcharsother{\let\do\@makeother \do\$\do\&\do\#\do\^\do\_\do\%\do\~}
\def\mn@doi{\begingroup\mn@urlcharsother \@ifnextchar [ {\mn@doi@}
  {\mn@doi@[]}}
\def\mn@doi@[#1]#2{\def\@tempa{#1}\ifx\@tempa\@empty \href
  {http://dx.doi.org/#2} {doi:#2}\else \href {http://dx.doi.org/#2} {#1}\fi
  \endgroup}
\def\mn@eprint#1#2{\mn@eprint@#1:#2::\@nil}
\def\mn@eprint@arXiv#1{\href {http://arxiv.org/abs/#1} {{\tt arXiv:#1}}}
\def\mn@eprint@dblp#1{\href {http://dblp.uni-trier.de/rec/bibtex/#1.xml}
  {dblp:#1}}
\def\mn@eprint@#1:#2:#3:#4\@nil{\def\@tempa {#1}\def\@tempb {#2}\def\@tempc
  {#3}\ifx \@tempc \@empty \let \@tempc \@tempb \let \@tempb \@tempa \fi \ifx
  \@tempb \@empty \def\@tempb {arXiv}\fi \@ifundefined
  {mn@eprint@\@tempb}{\@tempb:\@tempc}{\expandafter \expandafter \csname
  mn@eprint@\@tempb\endcsname \expandafter{\@tempc}}}

\bibitem[\protect\citeauthoryear{{Agarwal} et~al.,}{{Agarwal}
  et~al.}{2016}]{Agarwal2016}
{Agarwal} J.,  et~al., 2016, \mn@doi [\mnras] {10.1093/mnras/stw2179}, \href
  {https://ui.adsabs.harvard.edu/abs/2016MNRAS.462S..78A} {462, S78}

\bibitem[\protect\citeauthoryear{{Bertini} et~al.,}{{Bertini}
  et~al.}{2017}]{Bertini2017}
{Bertini} I.,  et~al., 2017, \mn@doi [\mnras] {10.1093/mnras/stx1850}, \href
  {https://ui.adsabs.harvard.edu/abs/2017MNRAS.469S.404B} {469, S404}

\bibitem[\protect\citeauthoryear{{Bird}}{{Bird}}{1994}]{1994mgdd.book.....B}
{Bird} G.~A.,  1994, {Molecular Gas Dynamics And The Direct Simulation Of Gas
  Flows}

\bibitem[\protect\citeauthoryear{{Ciarniello} et~al.,}{{Ciarniello}
  et~al.}{2022}]{Ciarniello2022}
{Ciarniello} M.,  et~al., 2022, \mn@doi [Nature Astronomy]
  {10.1038/s41550-022-01625-y}, \href
  {https://ui.adsabs.harvard.edu/abs/2022NatAs...6..546C} {6, 546}

\bibitem[\protect\citeauthoryear{{Della Corte} et~al.,}{{Della Corte}
  et~al.}{2016}]{DellaCorte2016}
{Della Corte} V.,  et~al., 2016, \mn@doi [\mnras] {10.1093/mnras/stw2529},
  \href {https://ui.adsabs.harvard.edu/abs/2016MNRAS.462S.210D} {462, S210}

\bibitem[\protect\citeauthoryear{{Della Corte} et~al.,}{{Della Corte}
  et~al.}{2019}]{DellaCorte2019}
{Della Corte} V.,  et~al., 2019, \mn@doi [\aap] {10.1051/0004-6361/201834912},
  \href {https://ui.adsabs.harvard.edu/abs/2019A&A...630A..25D} {630, A25}

\bibitem[\protect\citeauthoryear{{Drolshagen} et~al.,}{{Drolshagen}
  et~al.}{2017}]{Drolshagen2017}
{Drolshagen} E.,  et~al., 2017, \mn@doi [\planss] {10.1016/j.pss.2017.04.018},
  \href {https://ui.adsabs.harvard.edu/abs/2017P&SS..143..256D} {143, 256}

\bibitem[\protect\citeauthoryear{Duda \& Hart}{Duda \& Hart}{1972}]{Duda1972}
Duda R.~O.,  Hart P.~E.,  1972, \mn@doi [Commun. ACM] {10.1145/361237.361242},
  15, 11–15

\bibitem[\protect\citeauthoryear{{Fornasier} et~al.,}{{Fornasier}
  et~al.}{2015}]{Fornasier2015}
{Fornasier} S.,  et~al., 2015, \mn@doi [\aap] {10.1051/0004-6361/201525901},
  \href {https://ui.adsabs.harvard.edu/abs/2015A&A...583A..30F} {583, A30}

\bibitem[\protect\citeauthoryear{{Frattin} et~al.,}{{Frattin}
  et~al.}{2017}]{Frattin2017}
{Frattin} E.,  et~al., 2017, \mn@doi [\mnras] {10.1093/mnras/stx1395}, \href
  {https://ui.adsabs.harvard.edu/abs/2017MNRAS.469S.195F} {469, S195}

\bibitem[\protect\citeauthoryear{{Frattin} et~al.,}{{Frattin}
  et~al.}{2021}]{Frattin2021}
{Frattin} E.,  et~al., 2021, \mn@doi [\mnras] {10.1093/mnras/stab1152}, \href
  {https://ui.adsabs.harvard.edu/abs/2021MNRAS.504.4687F} {504, 4687}

\bibitem[\protect\citeauthoryear{{Fulle} et~al.,}{{Fulle}
  et~al.}{2016}]{Fulle2016}
{Fulle} M.,  et~al., 2016, \mn@doi [\apj] {10.3847/0004-637X/821/1/19}, \href
  {https://ui.adsabs.harvard.edu/abs/2016ApJ...821...19F} {821, 19}

\bibitem[\protect\citeauthoryear{{Fulle} et~al.,}{{Fulle}
  et~al.}{2018}]{Fulle2018}
{Fulle} M.,  et~al., 2018, \mn@doi [\mnras] {10.1093/mnras/sty464}, \href
  {https://ui.adsabs.harvard.edu/abs/2018MNRAS.476.2835F} {476, 2835}

\bibitem[\protect\citeauthoryear{{Fulle}, {Blum}, {Rotundi}, {Gundlach},
  {G{\"u}ttler}  \& {Zakharov}}{{Fulle} et~al.}{2020}]{Fulle2020}
{Fulle} M.,  {Blum} J.,  {Rotundi} A.,  {Gundlach} B.,  {G{\"u}ttler} C.,
  {Zakharov} V.,  2020, \mn@doi [\mnras] {10.1093/mnras/staa508}, \href
  {https://ui.adsabs.harvard.edu/abs/2020MNRAS.493.4039F} {493, 4039}

\bibitem[\protect\citeauthoryear{{Gerig} et~al.,}{{Gerig}
  et~al.}{2018}]{Gerig2018}
{Gerig} S.~B.,  et~al., 2018, \mn@doi [\icarus] {10.1016/j.icarus.2018.03.010},
  \href {https://ui.adsabs.harvard.edu/abs/2018Icar..311....1G} {311, 1}

\bibitem[\protect\citeauthoryear{{Gundlach}, {Fulle}  \& {Blum}}{{Gundlach}
  et~al.}{2020}]{Gundlach2020}
{Gundlach} B.,  {Fulle} M.,   {Blum} J.,  2020, \mn@doi [\mnras]
  {10.1093/mnras/staa449}, \href
  {https://ui.adsabs.harvard.edu/abs/2020MNRAS.493.3690G} {493, 3690}

\bibitem[\protect\citeauthoryear{{G{\"u}ttler} et~al.,}{{G{\"u}ttler}
  et~al.}{2017}]{Guttler2017}
{G{\"u}ttler} C.,  et~al., 2017, \mn@doi [\mnras] {10.1093/mnras/stx1692},
  \href {https://ui.adsabs.harvard.edu/abs/2017MNRAS.469S.312G} {469, S312}

\bibitem[\protect\citeauthoryear{{Hasselmann} et~al.,}{{Hasselmann}
  et~al.}{2021}]{Hasselmann2021}
{Hasselmann} P.~H.,  et~al., 2021, \mn@doi [\icarus]
  {10.1016/j.icarus.2020.114106}, 357, 114106

\bibitem[\protect\citeauthoryear{Hough}{Hough}{1962}]{hough1962method}
Hough P.~V.,  1962, Method and means for recognizing complex patterns

\bibitem[\protect\citeauthoryear{{Keller} et~al.,}{{Keller}
  et~al.}{2007}]{Keller2007}
{Keller} H.~U.,  et~al., 2007, \mn@doi [\ssr] {10.1007/s11214-006-9128-4},
  \href {https://ui.adsabs.harvard.edu/abs/2007SSRv..128..433K} {128, 433}

\bibitem[\protect\citeauthoryear{{Kelley}, {Lindler}, {Bodewits}, {A'Hearn},
  {Lisse}, {Kolokolova}, {Kissel}  \& {Hermalyn}}{{Kelley}
  et~al.}{2013}]{Kelley2013}
{Kelley} M.~S.,  {Lindler} D.~J.,  {Bodewits} D.,  {A'Hearn} M.~F.,  {Lisse}
  C.~M.,  {Kolokolova} L.,  {Kissel} J.,   {Hermalyn} B.,  2013, \mn@doi
  [\icarus] {10.1016/j.icarus.2012.09.037}, \href
  {https://ui.adsabs.harvard.edu/abs/2013Icar..222..634K} {222, 634}

\bibitem[\protect\citeauthoryear{{Kwon}, {Bagnulo}, {Markkanen}, {Agarwal},
  {Kolokolova}, {Levasseur-Regourd}, {Snodgrass}  \& {Tozzi}}{{Kwon}
  et~al.}{2022}]{Kwon2022}
{Kwon} Y.~G.,  {Bagnulo} S.,  {Markkanen} J.,  {Agarwal} J.,  {Kolokolova} L.,
  {Levasseur-Regourd} A.-C.,  {Snodgrass} C.,   {Tozzi} G.~P.,  2022, \mn@doi
  [\aap] {10.1051/0004-6361/202141865}, \href
  {https://ui.adsabs.harvard.edu/abs/2022A&A...657A..40K} {657, A40}

\bibitem[\protect\citeauthoryear{{Longobardo} et~al.,}{{Longobardo}
  et~al.}{2019}]{Longobardo2019}
{Longobardo} A.,  et~al., 2019, \mn@doi [\mnras] {10.1093/mnras/sty3244}, \href
  {https://ui.adsabs.harvard.edu/abs/2019MNRAS.483.2165L} {483, 2165}

\bibitem[\protect\citeauthoryear{{Longobardo} et~al.,}{{Longobardo}
  et~al.}{2020}]{Longobardo2020}
{Longobardo} A.,  et~al., 2020, \mn@doi [\mnras] {10.1093/mnras/staa1464},
  \href {https://ui.adsabs.harvard.edu/abs/2020MNRAS.496..125L} {496, 125}

\bibitem[\protect\citeauthoryear{{Longobardo} et~al.,}{{Longobardo}
  et~al.}{2022}]{Longobardo2022}
{Longobardo} A.,  et~al., 2022, \mn@doi [\mnras] {10.1093/mnras/stac2544},
  \href {https://ui.adsabs.harvard.edu/abs/2022MNRAS.516.5611L} {516, 5611}

\bibitem[\protect\citeauthoryear{{Lumme} \& {Bowell}}{{Lumme} \&
  {Bowell}}{1981a}]{Lumme1981I}
{Lumme} K.,  {Bowell} E.,  1981a, \mn@doi [\aj] {10.1086/113054}, \href
  {https://ui.adsabs.harvard.edu/abs/1981AJ.....86.1694L} {86, 1694}

\bibitem[\protect\citeauthoryear{{Lumme} \& {Bowell}}{{Lumme} \&
  {Bowell}}{1981b}]{Lumme1981II}
{Lumme} K.,  {Bowell} E.,  1981b, \mn@doi [\aj] {10.1086/113055}, \href
  {https://ui.adsabs.harvard.edu/abs/1981AJ.....86.1705L} {86, 1705}

\bibitem[\protect\citeauthoryear{{Mannel} et~al.,}{{Mannel}
  et~al.}{2019}]{Mannel2019}
{Mannel} T.,  et~al., 2019, \mn@doi [\aap] {10.1051/0004-6361/201834851}, \href
  {https://ui.adsabs.harvard.edu/abs/2019A&A...630A..26M} {630, A26}

\bibitem[\protect\citeauthoryear{{Merouane} et~al.,}{{Merouane}
  et~al.}{2017}]{Merouane2017}
{Merouane} S.,  et~al., 2017, \mn@doi [\mnras] {10.1093/mnras/stx2018}, \href
  {https://ui.adsabs.harvard.edu/abs/2017MNRAS.469S.459M} {469, S459}

\bibitem[\protect\citeauthoryear{{Mignone}, {Bodo}, {Massaglia}, {Matsakos},
  {Tesileanu}, {Zanni}  \& {Ferrari}}{{Mignone} et~al.}{2007}]{Mignone2007}
{Mignone} A.,  {Bodo} G.,  {Massaglia} S.,  {Matsakos} T.,  {Tesileanu} O.,
  {Zanni} C.,   {Ferrari} A.,  2007, \mn@doi [\apjs] {10.1086/513316}, \href
  {https://ui.adsabs.harvard.edu/abs/2007ApJS..170..228M} {170, 228}

\bibitem[\protect\citeauthoryear{{Ott} et~al.,}{{Ott} et~al.}{2017}]{Ott2017}
{Ott} T.,  et~al., 2017, \mn@doi [\mnras] {10.1093/mnras/stx1419}, \href
  {https://ui.adsabs.harvard.edu/abs/2017MNRAS.469S.276O} {469, S276}

\bibitem[\protect\citeauthoryear{{Pfeifer}, {Agarwal}  \&
  {Schr{\"o}ter}}{{Pfeifer} et~al.}{2022}]{Pfeifer2022}
{Pfeifer} M.,  {Agarwal} J.,   {Schr{\"o}ter} M.,  2022, \mn@doi [\aap]
  {10.1051/0004-6361/202141953}, \href
  {https://ui.adsabs.harvard.edu/abs/2022A&A...659A.171P} {659, A171}

\bibitem[\protect\citeauthoryear{{Rotundi} et~al.,}{{Rotundi}
  et~al.}{2015}]{Rotundi2015}
{Rotundi} A.,  et~al., 2015, \mn@doi [Science] {10.1126/science.aaa3905}, \href
  {https://ui.adsabs.harvard.edu/abs/2015Sci...347a3905R} {347, aaa3905}

\bibitem[\protect\citeauthoryear{{Tubiana} et~al.,}{{Tubiana}
  et~al.}{2015}]{Tubiana2015}
{Tubiana} C.,  et~al., 2015, \mn@doi [\aap] {10.1051/0004-6361/201525985},
  \href {https://ui.adsabs.harvard.edu/abs/2015A&A...583A..46T} {583, A46}

\bibitem[\protect\citeauthoryear{{Zakharov}, {Ivanovski}, {Crifo}, {Della
  Corte}, {Rotundi}  \& {Fulle}}{{Zakharov} et~al.}{2018}]{Zakharov2018}
{Zakharov} V.~V.,  {Ivanovski} S.~L.,  {Crifo} J.~F.,  {Della Corte} V.,
  {Rotundi} A.,   {Fulle} M.,  2018, \mn@doi [\icarus]
  {10.1016/j.icarus.2018.04.030}, \href
  {https://ui.adsabs.harvard.edu/abs/2018Icar..312..121Z} {312, 121}

\bibitem[\protect\citeauthoryear{{Zakharov}, {Rodionov}, {Fulle}, {Ivanovski},
  {Bykov}, {Della Corte}  \& {Rotundi}}{{Zakharov} et~al.}{2021}]{Zakharov2021}
{Zakharov} V.~V.,  {Rodionov} A.~V.,  {Fulle} M.,  {Ivanovski} S.~L.,  {Bykov}
  N.~Y.,  {Della Corte} V.,   {Rotundi} A.,  2021, \mn@doi [\icarus]
  {10.1016/j.icarus.2020.114091}, \href
  {https://ui.adsabs.harvard.edu/abs/2021Icar..35414091Z} {354, 114091}

\makeatother
\end{thebibliography}

%\begin{thebibliography}{99}
%\end{thebibliography}

%%%%%%%%%%%%%%%%%%%%%%%%%%%%%%%%%%%%%%%%%%%%%%%%%%

% Don't change these lines
\bsp	% typesetting comment
\label{lastpage}
\end{document}